\documentclass[sigconf]{acmart}

\usepackage{booktabs} 
\usepackage{tikz}
\usetikzlibrary{shapes,arrows}
\usepackage{pgfplots}
\usepackage{url}
\usepackage{epsfig}
\usepackage{amsmath}
\usepackage[linesnumbered,ruled]{algorithm2e}
\usepackage{graphicx}
\usepackage{array}
\usepackage{subcaption,siunitx,booktabs}
\newcolumntype{P}[1]{>{\centering\arraybackslash}p{#1}}

\newtheorem{defn}{\bf Definition}
\newtheorem{theorem}{{\bf Theorem}}

\setcopyright{rightsretained}

\acmDOI{10.475/123_4}

\acmISBN{123-4567-24-567/08/06}

\acmConference[WOODSTOCK'97]{ACM Woodstock conference}{July 1997}{El
  Paso, Texas USA}
\acmYear{1997}
\copyrightyear{2016}

\acmArticle{4}
\acmPrice{15.00}

\editor{Jennifer B. Sartor}
\editor{Theo D'Hondt}
\editor{Wolfgang De Meuter}

\begin{document}
\title{Using Randomness to Improve Robustness of Machine-Learning Models Against Evasion Attacks}
\titlenote{Produces the permission block, and
  copyright information}

 \author{Fan Yang}
 \affiliation{%
   \institution{University of Maryland, Baltimore County}
   \streetaddress{1000 Hilltop Circle}
   \city{Baltimore}
   \state{Maryland}
   \postcode{21250}
 }
 \email{fyang1@umbc.edu}

 \author{Zhiyuan Chen}
 \affiliation{%
   \institution{University of Maryland, Baltimore County}
   \streetaddress{1000 Hilltop Circle}
   \city{Baltimore}
   \state{Maryland}
   \postcode{21250}
 }
 \email{zhchen@umbc.edu}







\begin{abstract}
Machine learning models have been widely used in security applications such as intrusion detection, spam filtering, and virus or malware detection. 
However, it is well-known that adversaries are always trying to adapt their attacks to evade detection. 
For example, an email spammer may guess what features spam detection models use and modify or remove those features to avoid detection. 
There has been some work on making machine learning models more robust to such attacks. However, one simple but promising approach called {\em randomization} is underexplored. This paper proposes a novel randomization-based approach to improve robustness of machine learning models against evasion attacks. 
The proposed approach incorporates randomization into both model training time and model application time (meaning when the model is used to detect attacks). 
We also apply this approach to random forest, an existing ML method which already has some degree of randomness. Experiments on intrusion detection and spam filtering data show that our approach further improves robustness of random-forest method. We also discuss how this approach can be applied to other ML models.  
\end{abstract}

%
%



\maketitle

\section{Introduction}
With the arrival of big data era, data mining techniques have been widely used to build models for cyber security applications such as spam filtering \cite{blanzieri2008survey,cormack2007email}, malware or virus detection \cite{ye2008intelligent,christodorescu2008mining,siddiqui2008survey}, and intrusion detection \cite{lee1999data,chandola2009anomaly,axelsson2000intrusion,portnoy2001intrusion}. However, attackers may use a type of attacking strategy called {\em evasion attack} which modifies their data to avoid detection. For example, an email spammer may modify spam emails to drop certain words or symbols and a hacker can modify the signature of a malware or virus.




%

There has been studies on vulnerability of AI/ML models \cite{lowd2005adversarial}, especially more recently on deep learning models \cite{EvtimovEFKLPRS17,papernot2016limitations,grosse2016adversarial,Sharif:2016:ACR:2976749.2978392,gu2017badnets}. There also has been work on building more robust mining models against evasion attacks \cite{lowd2005adversarial,stone1998towards,liu2009game,barreno2006can,uther1997adversarial,bruckner2012static,kantarcioglu2008game,kantarciouglu2011classifier,biggio2009multiple,liu2010mining, ramoni2001robust,xu1998robust,globerson2006nightmare,yuan2009robust,kolcz2009feature, 
Goodfellow15, papernot2016distillation,lee2017generative}. Most existing work use deterministic models. However according to the Minimum Description Length (MDL) principle \cite{grunwald2007minimum}, a deterministic model should be concise to avoid overfitting the training data. 
However a concise model often uses a small number of features or a small number of features may have much larger impact on the output than other features. Attackers
can modify such features to evade detection. 
To understand the problem of deterministic models, let us look at the following example.

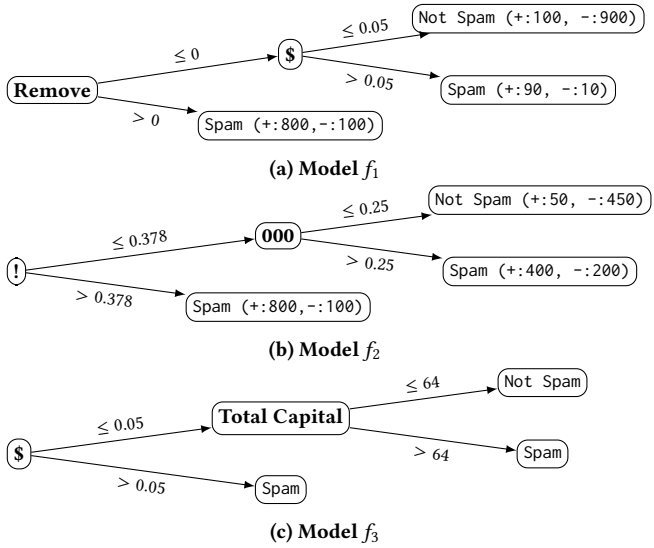
\begin{figure}[h]

\tikzset{
  treenode/.style = {shape=rectangle, rounded corners,
                     draw, align=center, scale=0.8
},
  root/.style     = {treenode  
},
  highlight/.style      = {treenode, font=\large\bfseries, 
},
  env/.style      = {treenode, font=\ttfamily\normalsize},
  dummy/.style    = {circle,draw}
}

\centering 
\begin{subfigure}[]{1.0\columnwidth}

\begin{tikzpicture}
  [
    grow                    = right,
    sibling distance        = 3em,
    level distance          = 10em,
    edge from parent/.style = {draw, -latex},
    every node/.style       = {font=\footnotesize},
    sloped
  ]
  \node [highlight] {Remove}
    child { node [env] {Spam (+:800,-:100)}
      edge from parent node [below] {$>$ 0} }
    child { node [highlight] {\$}
      child { node [env] {Spam (+:90, -:10)}
        edge from parent node [below] {$>$ 0.05} }
      child { node [env] {Not Spam (+:100, -:900)}
              edge from parent node [above, align=center]
                {$\leq$ 0.05}
              node [below] {}}
              edge from parent node [above] {$\leq$ 0} };

\end{tikzpicture}

\caption{Model $f_1$} \label{fig1:a}  
\end{subfigure}
\hfill
\begin{subfigure}[]{1.0\linewidth}

\begin{tikzpicture}
  [
    grow                    = right,
    sibling distance        = 3em,
    level distance          = 11em,
    edge from parent/.style = {draw, -latex},
    every node/.style       = {font=\footnotesize},
    sloped
  ]
  \node [highlight] {!}
    child { node [env] {Spam (+:800,-:100)}
      edge from parent node [below] {$>$ 0.378} }
    child { node [highlight] {000}
      child { node [env] {Spam (+:400, -:200)}
        edge from parent node [below] {$>$ 0.25} }
      child { node [env] {Not Spam (+:50, -:450)}
              edge from parent node [above, align=center]
                {$\leq$ 0.25}
              node [below] {}}
              edge from parent node [above] {$\leq$ 0.378} };

\end{tikzpicture}

\caption{Model $f_2$} \label{fig1:b}  
\end{subfigure}
\hfill
\begin{subfigure}[]{1.0\linewidth}

\begin{tikzpicture}
  [
    grow                    = right,
    sibling distance        = 3em,
    level distance          = 11em,
    edge from parent/.style = {draw, -latex},
    every node/.style       = {font=\footnotesize},
    sloped
  ]
  \node [highlight] {\$}
    child { node [env] {Spam}
      edge from parent node [below] {$>$ 0.05} }
    child { node [highlight] {Total Capital}
      child { node [env] {Spam}
        edge from parent node [below] {$>$ 64} }
      child { node [env] {Not Spam}
              edge from parent node [above, align=center]
                {$\leq$ 64}
              node [below] {}}
              edge from parent node [above] {$\leq$ 0.05} };

\end{tikzpicture}

\caption{Model $f_3$} \label{fig1:c}  
\end{subfigure}
    \caption{Decision tree models created by our model in Example 1}
      \label{fig:1}
    \end{figure}

\vspace{0.03in}
\noindent
{\bf Example 1:} Let us consider an email spam filtering example. Suppose the filtering software uses decision trees. Figure \ref{fig:1} shows several sample decision trees (in practice decision trees will have more nodes but here we simplify the trees to show the concept). 
Each node represents the fraction of words in an email that contain a word or symbol except for ``total capital'', which represents the total number of capitalized letters in the email. 

If only one decision tree, say $f_1$ is used, it is very easy for attackers to modify a spam email to evade detection. 
For example, suppose attackers have an email with feature values shown in Table \ref{table:1}(a). Attackers may learn that the spam filtering software looks at the ``remove'' and ``\$'' features. As a result attackers just need to modify one feature (the percentage of \$ sign in the email) to avoid detection. 


\begin{table}[htb]
  \centering
\begin{subtable}{0.5\textwidth}
\centering
  \begin{tabular}{|P{1.5cm}|P{0.5cm}|P{0.8cm}|P{1.5cm}|P{2.2cm}|}
    \hline
    Remove &  \$  & ! & 000 & Total Capital \\ \hline
    0    & 0.2 & 0.4 &  0.3 & 100   \\ \hline
  \end{tabular}
   \caption{Attacker's original spam email}
   \label{table1:a}
\end{subtable}
\bigskip

\begin{subtable}{0.5\textwidth}
\centering
  \begin{tabular}{|P{1.5cm}|P{0.5cm}|P{0.8cm}|P{1.5cm}|P{2.2cm}|}
    \hline
    Remove &  \$  & ! & 000 & Total Capital \\ \hline
    0    & \textbf{0.05} & 0.4 &  0.3 & 100   \\ \hline
  \end{tabular}
   \caption{Modified email to trick $f_1$ (bold as changes)}
   \label{table1:b}
\end{subtable}
\bigskip

\begin{subtable}{0.5\textwidth}
\centering
  \begin{tabular}{|P{1.5cm}|P{0.5cm}|P{0.8cm}|P{1.5cm}|P{2.2cm}|}
    \hline
    Remove &  \$  & ! & 000 & Total Capital \\ \hline
    0    & \textbf{0.05} & 0.4 &  0.3 & \textbf{64}   \\ \hline
  \end{tabular}
   \caption{Modified eamil to trick at least two models in Figure 1}
   \label{table1:c}
\end{subtable}
\bigskip

\begin{subtable}{0.5\textwidth}
\centering
  \begin{tabular}{|P{1.5cm}|P{0.5cm}|P{0.8cm}|P{1.5cm}|P{2.2cm}|}
    \hline
    Remove &  \$  & ! & 000 & Total Capital \\ \hline
    0    & \textbf{0.05} & \textbf{0.378} &  \textbf{0.25} & 100   \\ \hline
  \end{tabular}
   \caption{Modified email to trick $f_1$ and $f_2$}
   \label{table1:d}
\end{subtable}

  \caption{Some possible attacks for Example 1}
\label{table:1}
\end{table}

Some researchers try to use an ensemble approach (i.e., build a number of models instead of one) \cite{biggio2008adversarial,biggio2010multiple} to improve robustness of models. 
However, although ensemble approach does add some uncertainty to the generated models, 
it has two shortcomings: 1) its goal is still mining quality so the models built by ensemble approach may still frequently use a small subset of features, making it vulnerable to evasion attacks; 2) ensemble approach is still deterministic at model application time, making it easier for attackers to adapt.


\begin{figure}[tb]
\centering
\begin{tikzpicture}[scale=0.8]
\begin{axis}[
line width=0.2mm,
mark size=0.8,
xlabel=Features,
ylabel=Number of Trees using Each Feature]
\centering

\addplot[smooth,color=black,mark=o]
plot coordinates {
(1,91) 
(2,88)
(3,86)
(4,84)
(5,83)
(6,83)
(7,81)
(8,78)
(9,76)
(10,74)
(11,70)
(12,69)
(13,66)
(14,62)
(15,60)
(16,59)
(17,57)
(18,57)
(19,52)
(20,52)
(21,51)
(22,45)
(23,45)
(24,43)
(25,43)
(26,39)
(27,39)
(28,38)
(29,38)
(30,36)
(31,34)
(32,33)
(33,32)
(34,30)
(35,28)
(36,28)
(37,27)
(38,27)
(39,26)
(40,23)
(41,22)
(42,18)
(43,18)
(44,18)
(45,18)
(46,15)
(47,15)
(48,14)
(49,13)
(50,9)
(51,8)
(52,8)
(53,8)
(54,2)
(55,1)
(56,0)

};

\end{axis}

\end{tikzpicture}

    \caption{Number of trees using each feature by random forest on Spambase data set}
      \label{fig:tree-count}
    \end{figure}
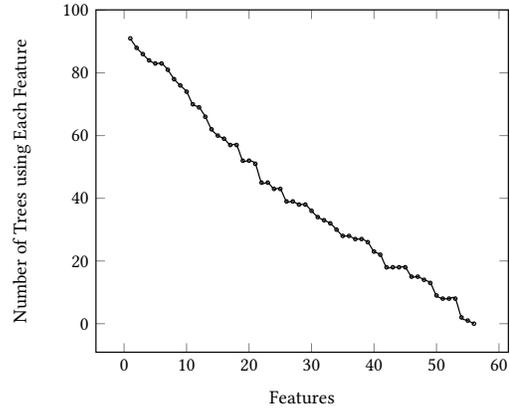

To validate the first shortcoming, we ran random forest (an ensemble method) on Spambase data set \cite{hettich-uci}, which is an email spam data set with  57 features and built 100 decision trees. 
Figure \ref{fig:tree-count} reports for each feature (attribute), the number of trees using that feature. 
9 out of 57 features appear in at least 80\% of trees and 22 features appear in at least half of trees. 
This means that attackers can modify these frequently used features to change prediction outcome. 



\vspace{0.03in}
\noindent
{\bf Our contributions:}
To address these shortcomings, our approach uses randomness to increase the uncertainty of models but at the same time does not violate the MDL principle for each individual model. We make the following contributions: 
\begin{enumerate}
\item
Methods to inject randomness into model building time by building a diverse pool of mining models. These methods are similar to ensemble learning but optimize the tradeoff between mining quality and robustness. These methods also require very little modification to existing algorithms. 
In this paper we used random forest as an example and showed how to modify the training algorithm for random forest to build a more robust model pool. 
We selected random forest because it already has some degree of randomness at model building time. However it does not optimize the tradeoff between robustness and accuracy of models. At the end of the paper we will discuss possible ways to extend our approach to other mining methods. 
\item
Methods to randomly select a subset of models at model application time (when the model is used for detection) to further boost robustness. 
\item
A theoretical framework that bounds the minimal number of features an attacker needs to modify given a set of selected models. 
\end{enumerate}

In Example 1, suppose our method builds a pool of three decision trees shown in Figure \ref{fig:1}. 
Table \ref{table:1} (c) shows the number of features an attacker needs to modify if all three trees are used at model application time (i.e., at least two trees need to return ``not spam''). Attackers need to modify at least two features now (using $f_1$ alone just needs one modification). 

Our method further boosts robustness by adding randomness at model application time. For example, 
if all three trees are used in Figure \ref{fig:1}.
As shown in Table \ref{table:1} (c), attackers only need to modify two features to avoid detection. 
However if we select $f_1$ and $f_2$ (or $f_2$ and $f_3$) at model application time and the spam filtering software will only label the email ``not spam'' if both trees return ``not spam'', attackers have to modify at least three features as shown in Table \ref{table:1} (d).

The rest of the paper is organized as follows. Section \ref{sec:related} describes related work. Section \ref{sec:background} introduces some background
information and gives an overview of our approach.
Section \ref{sec:random-forest} describes how our approach can be applied to random forest.
Section \ref{sec:framework} presents a theoretical framework to bound adversaries' cost. 
Section \ref{sec:experiment} presents experimental results. 
Section \ref{sec:extension} discusses how our approach can be extended to other mining models and Section \ref{sec:conclusion} concludes the paper.

\section{Related Work}
\label{sec:related}
Next we briefly describe related work in the literature.
The work can be roughly divided into five categories. 

The first category of work studies attacks against conventional mining models. Barreno et al. described a taxonomy of 
attacks and briefly mentions a few possible defense strategies 
\cite{barreno2006can}.
Lowd and Meek \cite{lowd2005adversarial}
studied an adversarial classifier reverse engineering algorithm to learn the models of a given learning algorithm. They assume that
the adversary can find out the outcome of a prediction model by sending instances to the model.
Attacks on intrusion detection are discussed in \cite{fogla2006polymorphic}.
Vulnerability of biometric systems are studied in \cite{uludag2004attacks,galbally2006vulnerability}.
Tramer et al. studied attacks that steal machine learning models based on output of such models \cite{tramer2016stealing}.

The second category of work tries to find robust learning algorithms in presence of attacks. A common approach is to model the learning problem as an optimization problem \cite{dekel2010learning,lanckriet2003robust,teo2007convex}. The gain of attackers is modeled as an additional term in the objective function along with the original objectives. This approach is especially suitable for linear classifiers because they have simple objective functions.
Globerson and Roweis \cite{globerson2006nightmare} studied the problem of building a robust SVM classifier in presence of feature deletion attacks (attackers can delete up to a certain number of features from the test data). 
The proposed method adds an additional term as the worst case accuracy loss caused by deletion attacks to the standard SVM's objective function and uses quadratic programming to solve the problem. Another adversarial learning method is proposed for SVM in \cite{zhou2012adversarial}, which considers two possible settings for attackers, in one of them the attackers
can corrupt data without any restriction and in the other one the attackers have costs associated with attacks.
Kolcz and Teo \cite{kolcz2009feature} proposed a simple feature weighting scheme for SVM and 
logistic regression.

The third category of work 
applies game theory to the adversarial learning problem. Typically the problem is modeled as a two-player and multi-stage game between the data miner and the adversary. At each stage, the adversary tries to find the best possible attacks and the data miner adjusts the mining models to such attacks. 
Kearns and Li \cite{kearns93siam} proposed a theoretical upper bound on tolerable malicious error rates. 
Dalvi et al. \cite{dalvikdd04} proposed a game 
theory framework which models the data miner and the adversary as a two player game. They assume that both players have perfect knowledge (i.e., data miner knows the adversary's attacking strategy and the adversary knows the mining model).  Several other works \cite{bruckner2011stackelberg,kantarciouglu2011classifier,bruckner2009nash,liu2010mining} model the adversarial learning problem as a Stackelberg game.

The fourth category of work focuses on vulnerabilities of deep neural nets \cite{EvtimovEFKLPRS17,papernot2016limitations,grosse2016adversarial,Sharif:2016:ACR:2976749.2978392,gu2017badnets}. Most of these studies focus on image classification and they have shown that adversarial examples can effectively fool a neural network to misclassify a slightly modified image. 
There has been some effort to make deep neural nets more robust \cite{Goodfellow15, papernot2016distillation,lee2017generative}, where most of them retrain the neural nets with added adversarial examples. However most of such work still focuses on images. Image-based methods operate in a
continuous feature space so they may not be directly applicable to cyber security applications which contain a lot of discrete attributes (e.g., words in emails or network protocol for intrusion detection data). To the best of our knowledge, the only exception is \cite{grosse2016adversarial}, where Grosse et al. have 
studied how to generate adversarial examples for malware classification and use these examples to retrain a deep neural net. 

Most of existing methods in these four categories use deterministic mining models. As we mentioned before, based on the MDL principle, deterministic models should be concise to avoid overfitting. This makes them vulnerable to attacks. We propose to use non-deterministic models to address this shortcoming. 

The final category of work use ensemble methods in adversarial settings \cite{ross2006handbook,hershkop2005combining,tran2008adjustable,perdisci2006using,skillicorn2009adversarial}.
For example, Biggio et al.  \cite{biggio2010multiple} used two methods: random subspace (a random subset of features are used for training each model) and bagging (a random sample of training data is used for training each model). 

Although these methods are similar to our approach, 
they do not consider the problem of optimizing the tradeoff between mining quality and robustness, which is the focus of our approach.
These solutions also only consider the model building time, and we will inject randomness into model application time as well.


The only work we are aware of that uses randomization at model application time is \cite{vorobeychik2014optimal}, where the authors considered an optimal strategy when 
the system can use several classifiers. They found the optimal solution is either to choose a classifier uniformly at random or choose the classifier with the smallest error depending on the relative importance between accuracy vs. robustness. However, this work does not consider how to create these classifiers and does not test their approach on real data sets. We propose a method to build more diverse pool of models and a clustering-based method to select these models at model application time. We also test our solution on real data sets. 

\section{Background and Overview of Our Approach}
\label{sec:background}

We first introduce some notations.
Let $\mathcal{L}$ be a data mining algorithm and $T=\{(x_i,y_i)\}_{i=1}^N$ be a set of N training samples 
where $x_i=(x_{i1},\ldots,x_{im})$ is a training instance with $m$ independent variables (or features) $A_1,\ldots,A_m$ and $x_{ij}$ is the value of $A_j$ in $x_i$, and $y_i$ is the value of dependent variable $Y$ for $x_i$. Let $M$ be the number of models in the model pool. 
We will build a pool  $\mathcal{P}=\{f_1(x), \ldots,$$f_M(x)\}$ of prediction models where each $f_i(x)$ is a model to predict the value of $Y$ for record $x$. 
At model application time, we will select a subset of $\mathcal{P}$ for prediction. 

\noindent
{\bf Threat Model:}
This paper focuses on evasion attacks where attackers can observe the prediction outcome of the detection model and modify their attacking instances accordingly to avoid detection, but cannot tamper with the model and training data directly (also called {\em exploratory integrity attacks} in \cite{barreno2006can}). 

Such attacks are quite common in practice. For example, Huang et al. \cite{huang2016android} found that many attackers are checking whether their malware will be detected by malware scanning software by uploading their malware programs on scanning sites such as VirusTotal and keep modifying their malware until they are not detected. 

We will consider two cases for attackers' knowledge. 

\begin{defn} 
Complete knowledge of model pool: Attackers know the models being built over training data and can modify their attacking instances based on their knowledge. 
\label{defn:complete}
\end{defn}

\begin{defn} 
Incomplete knowledge of model pool: Attackers do not know the models being built over training data, but can guess what features may be considered important. 
\label{defn:incomplete}
\end{defn}

The incomplete knowledge case is more common in practice. The complete knowledge case often occurs if attackers are insiders. 
 
\begin{defn} 
Attackers' cost function: 
attackers have a cost $c(x_i,x_i')$ of modifying an instance $x_i$ into $x_i'$.
In this paper we assume that the cost function is a weighted sum of cost of modifying each feature,
i.e., $c(x_i,x_i')=\sum_{j=1}^{m} c_j dist(x_{ij},x_{ij}')$ where $x_{ij}$ is the $j$-th feature of $x$, $dist$ is a distance measure between the $j$-th feature
of $x_i$ and $x_i'$ and $c_j$ is a weight.
\label{defn:cost}
\end{defn}

 For example, if $dist(x_{ij},x_{ij}')$ is squared distance, $c(x_i,x_i')$ is square of weighted Euclidean distance.
In cases such as email spam detection, attackers can drop or add certain words to their spam emails. 
So we can set $dist(x_{ij},x_{ij}')$ to be the indicator function, i.e., it equals one if $x_{ij} \neq x_{ij}'$ and zero otherwise. The cost is now sum of weights of modified features (weight is $c_j$). 
$c_j$ can be set using domain knowledge. For example, if a feature is difficult to modify or it is crucial for the gain of attackers, $c_j$ should be large.


\begin{figure}[tb]

\tikzstyle{block} = [rectangle, draw, fill=blue!20, 
    text width=10em, text centered, rounded corners, minimum height=4em]
\tikzstyle{data} = [trapezium, trapezium left angle=70, trapezium right angle=110, 
text width=5em, minimum height=1cm, draw=black, fill=blue!30]
\tikzstyle{line} = [draw, -latex']
\tikzstyle{init} = [draw, ellipse,fill=red!20, node distance=3cm,
    minimum height=2em]
    
\begin{tikzpicture}[node distance = 2cm, auto]
    \node [init] (init) {User};
    \node [block, below of=init, node distance=2cm] (selection) {Random Selection of Models at Model Application Time};
    \node [block, below of=selection, node distance=2cm] (pool) {Randomized Model Pool Creation};
    \node [block, below of=pool, node distance=2cm] (algorithm) {Data Mining Algorithm L};
    \node [data, left of=algorithm, node distance=3.5cm] (data) {Training Data T};
    \path [line] (init) -- node{Attacking Instances}(selection);
    \path [line] (selection) -- node{Results}(init);
    \path [line] (pool) -- node{Model Pool}(selection);
    \path [line] (data) -- node {}(pool);
    \path [line] (algorithm) -- node {}(pool);

\end{tikzpicture}

    \caption{Decision tree models created by our model in Example 1}
      \label{fig:architecture}
    \end{figure}
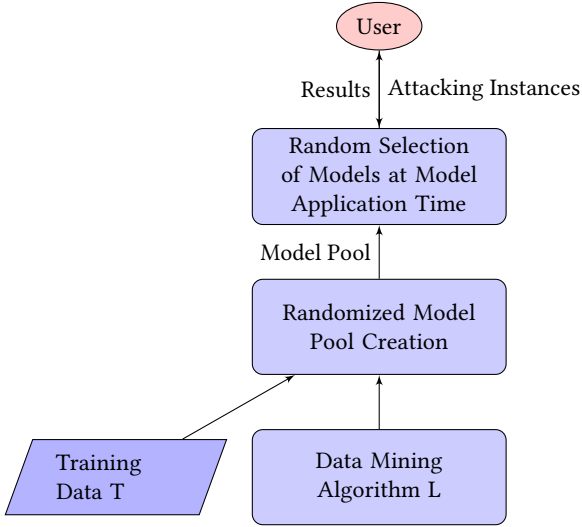

Figure \ref{fig:architecture} shows the architecture of our approach. 
Given a training data set $T$ and a data mining algorithm $L$, our approach first 
generates a model pool using randomization. When user provides an instance $z$ to classify, a  random selection process will be used to select a subset of models from the model pool to return a prediction for $z$. 
Next we show how our approach can be applied to random forest method. 

\section{Our Solution to Random-Forest}
\label{sec:random-forest}

Random forest method builds a pool of decision trees. Each decision tree is built on a random sample of training data (this is called {\em bagging}). When building the decision tree, a feature will be selected at each split based on some splitting criteria such as information gain. Unlike traditional decision tree building algorithms, random forest selects the splitting feature from a random subset of all features (this is called {\em random subspace}). 

Bagging and random subspace increase uncertainty at training time. However, random forest does not consider robustness of models against evasion attacks when selecting splitting feature. As shown in Figure \ref{fig:tree-count}, a few features appear in majority of trees built by random forest, making them easy target to evasion attack.
We developed a Weighted Random Forest algorithm (Section \ref{sec:weight}) to address this issue. 
The key idea is to penalize features with high vulnerability (or more frequent appearance) such that features are used more uniformly in different trees.
We also propose a clustering based method to add randomness at model application time (Section \ref{sec:clustering}).

\subsection{Weighted Random Forest Method}
\label{sec:weight}



\begin{algorithm}
    \SetKwInOut{Input}{Input}
    \SetKwInOut{Output}{Output}

    \Input{Training data set $T=\{(x_i,y_i)\}_{i=1}^N$,  number of models $M$, feature subset size $F$}
    \Output{A set of candidate models $f_1, \ldots, f_{M}$}
    Compute weights $w_1,\ldots,w_m$ for each feature

  \For {$i=1$ to $M$}
  {Draw a uniform random sample with replacement of size $N$ from $T$, let the sample be $T_i$
  
  Build a decision tree $f_i$ on $T_i$ where at each node a random subset of $F$ features are used and the splitting criteria for each feature $A_j$ is multiplied by weight $w_j$}

  \Return $f_1,\ldots,f_{M}$

    \caption{Weighted Random Forest Method}
    \label{algo:1}
\end{algorithm}

Algorithm \ref{algo:1} shows the pseudo code of the Weighted Random Forest algorithm. 
The algorithm is the same as random forest except that when the algorithm selects a feature to split, the splitting criteria of each feature $A_i$ is multiplied by a weight $w_i$. This weight will be used to optimize the tradeoff between mining quality and robustness. 

We compute the weights based on the following two observations.
First, features with higher modification cost should receive higher weights. 
Second, if modifying a feature is likely to change the classification outcome of a positive (malicious) instance to negative (benign), then this feature should have a lower weight. 
We call such features {\em vulnerable features} and use a metric called {\em differential ratio} to quantify vulnerability of a feature. 

\noindent
{\bf Differential Ratio:}
One observation is that a vulnerable feature is often very useful in distinguishing positive instances from negative instances. 
Thus measures such as information gain could be used to measure vulnerability.

However we found that such measures are mainly for classification accuracy, and they do not precisely capture vulnerability at times. 
So we proposed an alternative metric called {\em differential ratio} to quantify vulnerability. 
We start by considering binary trees, and then explain the difference between differential ratio and information gain, and finally generalize our solution to multi-branch trees.

Let $p_+(n_l)$ be the fraction of positive training instances in the subtree rooted at node $n$'s left child  and $p_+(n_r)$ be the fraction of positive instances in the subtree rooted at its right child. 
Let $|n|$ be the total number of training instances in the subtree rooted at node $n$ and $|T|$ be the total number of training instances. 
Let $A_n$ be the splitting feature used at node $n$.
We calculate a {\em differential ratio} for feature $A_n$ at $n$ as
\begin{equation}
d(A_n,n)=|p_+(n_l)-p_+(n_r)| \frac{|n|}{|T|}
\label{eq:dfN}
\end{equation}

Next we explain the intuition behind Equation \ref{eq:dfN}.
Let $x$ be a positive instance that falls under the subtree rooted at $n$. Modifying the splitting feature of $n$ may change $x's$ classification outcome from
positive to negative.
Now we try to estimate the probability of this change. We assume that test cases will follow similar distribution as the training cases (this is the basis for all 
data mining algorithms). Thus we can estimate the probability of a test case reaching node $n$ by $\frac{|n|}{|T|}$.
Since the test case is positive, it is more likely belonging to the child node with higher fraction of positive cases. 
Without loss of generality we assume that the left child has higher fraction of positive cases. 
Let $A_n$ be the splitting feature at node $n$. Modifying $A_n$ will move $x$ from left child to right child. We use $p_+(n_l)$ to approximate the probability of $x$ classified as positive in the left child, and $p_+(n_r)$ to approximate the same probability in right child. 
So the probability of $x$ being classified as negative after modifying $A_n$ can be estimated by
$p_+(n_l)(1-p_+(n_r))$. 

However this measure has two problems: 1) it is not symmetric; 2) it is greater than zero even if the right child has higher fraction of positive instances (so moving $x$ to right child will not help the attackers). So instead we use $|p_+(n_l)-p_+(n_r)|$ which is both symmetric and indicates that $x$ is more vulnerable when one of its child has much higher fraction of positive instances than the other child (so moving $x$ to the other child helps attackers). 

 %
\noindent
{\bf Difference with Information Gain:}
Information gain for node $n$ is defined in Equation \ref{eq:infor-gain}
\begin{equation}
IG(n)=H(n)-(\frac{|n_l|}{|n|}H(n_l)+\frac{|n_r|}{|n|}H(n_r))
\label{eq:infor-gain}
\end{equation}

Here $H(n)$, $H(n_l)$, and $H(n_r)$ are entropy for node $n$, its left child $n_l$ and right child $n_r$, respectively. 
There are two main differences between information gain and differential ratio: 1) information gain considers original class distribution ($H(n)$), 
differential ratio does not;
2) information gain considers size of each child node (the entropy of each child node is weighted by size), differential ratio does not. 

Figure~\ref{fig:1} shows the difference between differential ratio and information gain. The numbers in parenthesis are number of positive or negative instances in each node. 
The differential ratio for the ``\$'' node in tree $f_1$ equals $|0.1-0.9|\frac{1100}{2000}=0.88$, and the differential ratio for the ``000'' node in tree $f_2$ is $|0.1-0.67|\frac{1100}{2000}=0.62$. So the first one has higher differential ratio. 
However, if we compute information gain (IG), IG for ``\$'' node is 0.19 and IG for ``000'' node is 
0.26. So the second node has higher IG. The reason is that information gain considers entropy before the split as well as each child node's size, while differential ratio only considers the difference of fraction of positive instances in two child nodes. Here the ``\$'' node has one child with mostly
positive instances and the other with mostly negative instances, so modifying ``\$'' feature is more likely to change a positive instance to a negative instance.
On the other hand, the higher IG for ``000'' is mostly due to the higher entropy before the split, which is not directly related to vulnerability.

\noindent
{\bf Weighting Scheme:}
We then define a feature $A_j$'s differential ratio in a pool of trees as the sum of ratio in each tree divided by $M$ (total number of trees), where the ratio in each tree is the maximal ratio of all nodes having $A_j$ as splitting attribute, i.e., 
\begin{equation}
d(A_j, \mathcal{P}) = \frac{\sum_{f\in \mathcal{P}}\sum_{n \in f, n \ \mbox{\small splits on} \ A_j} \max{d(A_j,n)}}{M}
\label{eq:dfA}
\end{equation}

Here we take the maximal ratio in a tree so we consider the most vulnerable case for a feature $A_j$ (i.e., the worst case). 
Since this ratio depends on a model pool $\mathcal{P}$, we can run the original random forest (not weighted one) once to create a model pool $\mathcal{P}_0$
and calculate the ratio based on $\mathcal{P}_0$. 
We then compute weight $w_j$ for feature $A_j$ as
\begin{equation}
w(A_j) = e^{-r \frac{d(A_j,\mathcal{P}_0)}{c(A_j)}}
\label{eq:weight}
\end{equation}

Here $c(A_j)$ is the cost of modifying $A_j$ and $\mathcal{P}_0$ is the pool built by random forest. 
If we know all data instances ($x_i$) attackers have and their attacking strategy (i.e., $x_i'$), we can compute exact $c(A_j)$. However in practice
such information is not available. So we can approximate that cost with $c_j$ (i.e., the weight for $A_j$
in Definition \ref{defn:cost}).
$r$ is a parameter to adjust the importance of robustness.
If $r=0$, $w(A_j)=1$ for all features and our method is identical to random forest. For a positive $r$ value, 
the weight of a feature increases with the cost of modifying that feature and decreases with the differential ratio. So our method favors features that have higher
cost or are less vulnerable.  

The exponential function in Equation \ref{eq:weight} is used for smoothing. 
For example, if a feature $A_1$ has a differential ratio of
0.53 and a feature $A_2$ has a differential ratio of 0.01, suppose both features have cost of 1. Without the exponential function the weight of $A_2$ will be 53 times of that of $A_1$. This may penalize feature $A_1$ too much because features with high differential ratio are often features that can better distinguish positive instances from negative instances. With the exponential function the weight for $A_1$ is 0.45 and weight for $A_2$ is 0.98 when $r=1.5$. 
We will discuss how to choose appropriate value of $r$ in Section \ref{sec:experiment}.

\noindent
{\bf Generalization to multi-branch trees:} To generalize differential ratio to multiple-branch trees, we divide children of a node $n$ into two groups. The first group consists of child nodes with majority as positive training instances, and the second group consists of nodes with majority as negative training instances (if one of the groups is empty then differential ratio of $n$ is zero). 
Let $p_+(n_+)$ be the fraction of positive instances in the first group and $p_+(n_-)$ be the fraction of positive instances in the second group. We define differential ratio for feature $A_n$ at node $n$ as
\vspace{-0.1in}
\begin{equation}
d(A_n, n)=|p_+(n_+)-p_+(n_-)| \frac{|n|}{|T|}
\label{eq:df-multi}
\end{equation}
We then use this differential ratio in Equation \ref{eq:dfA} and \ref{eq:weight}.

\noindent
{\bf Computational complexity:}
The cost of random forest is $O(m M N \log N)$ where $m$ is number of features, $N$ is number of training instances and $M$ is number of models. WRF builds models twice (the first pass to generate $\mathcal{P}_0$ without the weighting scheme and the second pass  with the weighting scheme).
Once $\mathcal{P}_0$ is generated, computing differential ratio just needs to traverse each tree in $\mathcal{P}_0$ and costs $O(\max|f|M)$ where $\max|f|$ is the maximal number of nodes in a tree. Normally $N >> \max|f|$, so the complexity of WRF is $O(m M N \log N)$.  

\subsection{Clustering-based Model Selection at Model Application Stage}
\label{sec:clustering}

At model application stage we can dynamically select a subset of models for each test case such that it is even harder for attackers to find out what models are used. 
However, the mining quality is also very important and for ensemble methods such as random forest using too few models often leads to poor mining quality. So we need to balance robustness and mining quality at model application time as well. 

We propose a clustering-based model selection method (shown in Algorithm \ref{algo:2}). 
This method is based on the observation that if two trees share very few common features then they should be robust to evasion attacks because attackers need to modify more features. 

\begin{algorithm}
    \SetKwInOut{Input}{Input}
    \SetKwInOut{Output}{Output}

    \Input{A model pool $\mathcal{P}=\{f_1, \ldots, f_{M}\}$, parameters $s$,$q$, and a test case $t$}
    \Output{A subset of models to classify $t$}
    Creates a similarity graph $G=(V,E)$ where node $v\in V$ is a tree in $\mathcal{P}$ 
 and two nodes are linked by an edge $e$ if they share common features 
 and $e$'s weight is the sum of differential ratio of shared features

   Use spectral clustering to create $s$ clusters

At model application time, randomly select $q$ models per cluster and return them
    \caption{Clustering-based Model Selection Algorithm}
    \label{algo:2}
\end{algorithm}


The algorithm first creates a similarity graph where each node is a tree in $\mathcal{P}$ and two nodes are linked if they share common features and the weight of the link is the sum of differential ratio of shared features. It then uses spectral clustering to divide the models into $s$ clusters such that there are few between cluster links. The clustering step can be done offline. At model application time, 
for each test case, the clustering method randomly selects $q$ models from each cluster. 
These $qs$ models will be used to classify this test case. Note that different models will be used to classify different test cases. Since models in different clusters share very few common features, the selected models also share fewer common features than the original model pool. 
We will discuss how to empirically select $q$ and $s$ in Section \ref{sec:experiment}.

Let $M$ be the number of trees and $m$ be the number of features. It takes $O(m M^2)$ time to build the similarity graph. The cost of spectral clustering is $O(M^3)$. So the computational complexity of the clustering-based method is $O(M^3+m M^2)$. Note that this cost is not related to number of instances. 

The clustering-based algorithm can use models generated by our weighted (WRF) algorithm. We call the combined algorithm Cluster-based Weighted Random Forest (CWRF). 

\section{Theoretical Framework}
\label{sec:framework}

We propose a theoretical framework to provide a lower bound to attackers' effort. 
We will consider the case when the distance function $dist$ in Definition \ref{defn:cost} is indicator function because it is easier to reason with.
In addition, in many real life applications such as spam filtering, indicator function is appropriate. So the cost of modifying a set of features in a set
$SA$ equals $\sum_{A_j \in SA} c_j$. 
We start with a theoretical bound for random forest. In Appendix \ref{sec:extend-model} we will generalize it to other mining methods. 
We first introduce some notations. 

\begin{defn}
Let $f$ be a decision tree to detect whether data instance $x$ is positive or negative as a cyber security threat (e.g., a spam, an intrusion, or a fraud). 
A critical path of $f$ is a root-to-leaf path $p$ with nodes $n_1, n_2, \ldots, n_{|p|}$ where $n_{|p|}$ is a leaf node with positive label.
\end{defn}

It is clear that any instance labeled positive must lie on one of the critical paths (actually at the leaf node) and to turn such an instance into negative, 
attackers need to modify some features on the critical path.

\begin{defn}
Critical count $CC(A_j)$ for a feature $A_j$ in a set of models $\mathcal{P}=\{f_1,\ldots,f_M\}$ equals the number of trees in $\mathcal{P}$ that have $A_j$ on at least one critical path.
\end{defn}

For example, in Example 1, the critical count for ``\$'' is two because it appears in critical paths in $f_1$ and $f_3$. The critical count for 
the remaining features is all one because each only appears in one tree's critical paths. 
Next we give the bound. 

\begin{theorem} 
A pool $\mathcal{P}=\{f_1(x),$$ \ldots,$$f_M(x)\}$ of decision trees satisfies $(t_1,t_2,k)$-robustness if 
for any set $SA$ of $k$ features, $\sum_{A_j \in SA} CC(A_j) \leq t_1$ and 
total modification cost $\sum_{A_j \in SA} c_j \geq t_2$. 
So for any positive data instance $x$ with current positive vote (i.e., number of trees labeling $x$ as positive) greater or equal to $\lceil M/2 \rceil + t_1$, an attacker needs to modify more than $k$ features or pay more than $t_2$ modification cost to let $x$ avoid detection by $\mathcal{P}$.
\label{th:bound}
\end{theorem}

The proof is quite straightforward. Suppose $x$ is classified as positive by a tree $f$. To modify $x$ such that $f$ will classify $x$ as negative, an attacker must modify some feature on critical paths of $f$. Since $CC(A_i)$ is the number of trees with feature $A_i$ on their critical paths, modifying $A_i$
can reduce the positive vote count by at most $CC(A_i)$. Since the sum of any $k$ features' critical count is at most $t_1$, the change in positive vote count by modifying $k$ features is at most $t_1$. Since the current positive vote count is no less than $\lceil M/2 \rceil + t_1$, the new count is at least $\lceil M/2 \rceil$ after changing $k$ features. Thus $x$ will be still classified as positive.
Since we need to modify more than $k$ features to change the outcome, the modification cost is at least $t_2$ because changing $k$ features already costs at least $t_2$.




For example, let $\mathcal{P}=\{f_1,f_2\}$ in Figure \ref{fig:1}, the maximal critical count for each feature is one, so $\mathcal{P}$ satisfies $(1,c*,1)$-robustness where $c*$ is the minimal $c_j$ among those features. 
According to Theorem \ref{th:bound}, for any instance with positive vote 2 (e.g., the original instance shown in Table \ref{table:1} (a)), the attacker needs to change at least 2 features to avoid detection. 
This bound is also tight because for a test case with feature ``Remove'' $=0$, ``\$'' $=0.2$, ``!'' $=0.2$, and ``0000'' $=0.3$,
attackers just need to  change ``\$'' to 0.05 and ``000'' to 0.25 to change the prediction of both $f_1$ and $f_2$ from spam to not spam.

Theorem \ref{th:bound} provides a worst-case bound. However in practice the performance is usually better because not every modification of
a feature on a critical path will lead to change of classification outcome. For instance, for the first test case in Table \ref{table:1} (a), attackers need to modify three features instead of two for model set $\{f_1,f_2\}$ as shown in Table \ref{table:1} (d). The differential ratio proposed in Equation \ref{eq:dfA} can be seen as a more realistic estimation of robustness but it does not give worst case bound. 
%

\noindent
{\bf Worst-case bound after clustering:} The clustering-based model selection method also provides a worst-case bound. If each feature does not appear in more than $l$ clusters, then 
the critical count of each feature is no more than $lq$ because we only select $q$ models per cluster. 
So the total critical count of $k$ features will not exceed $klq$. The selected models satisfy $(klq, c*, k)$-robustness where $c*$ is the minimal sum of $c_j$ of $k$ features. 
For example, suppose in Figure \ref{fig:1} the trees are divided into two clusters, the first cluster with tree $f_1$ and $f_3$ and the second with
tree $f_2$. Each feature only shows up in one cluster, i.e., $l=1$. We also select one model per cluster so $q=1$. So the selected models (say $f_1$ and $f_2$) satisfy $(1,c*,1)$-robustness and by Theorem \ref{th:bound} attackers need to modify at least 2 features for any instance with two positive votes. 

\section{Experimental Results}
\label{sec:experiment}
This section presents experimental evaluation of our proposed methods. Section \ref{sec:setup} describes setup of our experiment.
Section \ref{sec:tuning} discusses how we tune parameters of proposed methods and 
Section \ref{sec:result} compares proposed methods to existing ones. 

\subsection{Setup}
\label{sec:setup}

\noindent
{\bf Algorithms:}
We compare the following algorithms:
\begin{enumerate}
\item
WRF: this is the proposed weighted algorithm in Algorithm \ref{algo:1} without the clustering step. 
\item
CWRF: this is the proposed algorithm with both feature weighting at model building time and cluster-based model selection (in Algorithm \ref{algo:2}) at model application time. 
\item
IG: this method is a variant of the proposed WRF method. The only difference is that IG replaces differential ratio with information gain in Equation \ref{eq:weight}.
Note that IG is different from random forest because features with high information gain are penalized by having lower weights according to Equation \ref{eq:weight}.
\item
RF: this is the original random forest algorithm. 
RF can be seen as a special case of WRF when $r=0$ (i.e., weights are uniform). 
\item
C4.5. This method builds a single decision tree model. 
\item
avg-LR: this method uses random subspace to build a number of logistic regression models and computes the average weight of each feature in these models as the weight for the overall model \cite{kolcz2009feature}. It was the method with the best performance as reported in \cite{kolcz2009feature}. This method injects randomness at model building time but differs from our method on three aspects: 1) it does not optimize tradeoff between accuracy and robustness; 2) it does not 
inject randomness in model application time; 3) it uses logistic regression rather than random forest. 
\end{enumerate}

\noindent
{\bf Data sets:}
We used the Spambase data set (an email spam data set) from UCI Machine Learning Repository \cite{hettich-uci} and  
network traffic data from Kyoto University's Honeypot (an intrusion detection data set) \cite{song2011statistical}. 
For the Kyoto University data set, we randomly selected a sample of 45,390 instances from data collected in December 2015.
Since the original data set is quite skewed (with mostly normal traffic), we under-sampled normal traffic data and kept about half sample normal and half attacks. 
We also removed duplicates from the data set. 
We call this sampled data set {\em Kyoto-Sample}. 
Details of these data sets can be found in Table \ref{tab:data}.
Spambase only contains numerical features. Kyoto University data has both numerical and categorical features.

Features for Spambase data are mainly frequency of words and symbols as well as length of sequence of capital letters. 
For Kyoto University data set we only used features extracted from raw traffic data such as duration of connection, number of bytes sent by source IP. 
Attackers can easily modify features in both data sets. E.g., attackers can add or drop a word or symbol or capital letter sequences in a spam email for Spambase and 
modify the number of bytes sent by source IP for Kyoto data set.

For Spambase, we randomly selected 70\% of data for training and the remaining for testing. 
For Kyoto-Sample, we randomly selected 10 days of data for training and the remaining for testing.


\begin{table}[htbp]
\centering

\begin{tabular}{|c|c|c|l|} 
\hline
Data set &	Number of  & 	Number of & Number \\
&instances&features&of positive \\
&&&instances \\
\hline
Spambase&4,601&57&	1813\\
\hline
Kyoto-Sample  &45,390&14& 22,687\\
\hline
\end{tabular}
\caption{Characteristics of data sets}
\label{tab:data}
\vspace{-0.1in}
\end{table}

\noindent
{\bf Attacking strategies:}
For simplicity we assume that all features have uniform modification cost and distance function between two feature values $dist(x_{ij},x_{ij}')$ to be the indicator function so attackers' effort will be minimized if they need to modify the smallest number of features. 

We simulated two simple but yet effective attacking strategies in our experiments: one for incomplete knowledge and the other for complete knowledge. 
These attacking strategies are not necessarily used in practice but we find they are quite effective against existing machine learning models.
So we use them to compare our proposed methods with existing methods. 

Both 
attacking strategies use a random probing phase to learn the optimal attacking order over features. 
For example, in Figure \ref{fig:1}, the optimal order is to modify the feature ``\$'' first because this may change the result of two out of three models. 
In attacking phase, attackers use the learned order to launch evasion attacks. 
 
We assume that attackers have obtained a set of instances for its probing and this set is called {\em probing set}.
A probing set should contain both positive (malicious) instances and negative (benign) instances. 
Positive instances are used for probing and negative instances are used to modify feature values. 
The probing set does not have to overlap with the training data used to build the machine learning models but it should follow similar distribution. 

For Spambase, we randomly divide test set into two halves. We assume that attackers will use the first half as probing set and the remaining 
half for launching attacks based on learned strategy. 
For Kyoto-Sample, we randomly selected 7 days of data in test set as probing set and the remaining to launch attacks.

In the probing phase, attackers randomly select a positive instance in its probing set and send it to the machine learning model and check whether  the mining model will reject the instance (i.e., labeled as positive). 

If the instance is rejected, attackers replace a randomly selected feature's value with the mean value of that feature in negative instances in the probing set. 
For example, in Figure \ref{fig:1}, attackers can reduce the number of ``\$'' in the spam email such that its frequency equals the frequency in non spam emails. 
The modified instance is resent to the mining model. This process continues until the instance is accepted. 
Attackers record the last modified feature $A_l$
 because classification outcome changes from positive to negative as a result of modifying $A_l$. 
Note that a feature $A_i$ modified before the last feature may contribute to the success of attack as well. However, we choose to be conservative here because we are uncertain about this. 

After probing a certain number of positive instances, attackers sort recorded features by their frequencies in descending order.
In both data sets, we varied number of probed positive instances from 50 to 200 and found little difference in results. This means attackers can learn an effective attacking strategy quite quickly. We report results when 50 positive instances are probed. 

In the attacking phase, attackers modify each positive instance in the attacking set one feature at a time in the order obtained after probing and stop when the modified instance is classified as negative. 
The attacking strategy for the complete knowledge case is the same as incomplete knowledge case except that attackers probe the whole set of positive test cases (including those used to launch actual attack) in the probing phase because attackers already know the machine learning model and can test them locally.
 
Interestingly, this random probing strategy is better than other strategies we tried (including 
a greedy strategy that selects features that lead to maximal decrease of positive votes for random forest), maybe because the former is less likely to get stuck in local maxima.

\noindent
{\bf Metrics:} We consider two cases when measuring robustness of a machine learning model. In the first case, attackers can modify up to $k$ features per instance where $k$ is a parameter. We call this case {\em bounded cost} because attackers' cost is bounded by $k$. We measure attackers' success rate as the percentage of modified positive instances used in the attacking phase that are eventually labeled negative by the machine learning model. Clearly, the lower the success rate, the more robust the model. 

In the second case, attackers can modify as many features as they want until each positive instance used in the attacking phase is labeled negative by the machine learning model. 
We call this case {\em unbounded cost} case. 
We compute average number of modified features in this case. The higher the average, the more robust the model. 

To measure mining quality, we used true positive rate and false positive rate. 
We found that in our experiments false positive rate for all algorithms in all settings is always below 10\%. So we only report results for true positive rate.
Since there is some randomness in mining algorithms and attacking strategies, we ran each experiment 20 times and took the average of results.

All experiments were run on a desktop computer with Intel $i7$ quad core processor, 32 GB RAM, 2 TB hard disk, and running Windows 7. 
All algorithms were implemented in Java by extending source code for Weka 3.8. 

\vspace{-0.1in}
\subsection{Tuning of Parameters}
\label{sec:tuning}

We need to set three parameters for our proposed WRF and CWRF methods:
 1) $r$ which is used in Equation \ref{eq:weight} to adjust the tradeoff between robustness and mining quality;
2) $s$ as the number of clusters for CWRF; 3) $q$ as the number of models selected at model application time from each cluster of models.

\begin{figure*}[!htb]
\begin{minipage}[t]{0.49\linewidth}
\centering
\begin{tikzpicture}[scale=0.7]
\begin{axis}[
ymin=0.5, ymax=1,
legend pos=south west,
line width=0.2mm,
mark size=2.5,
xlabel=$r$ value,
ylabel=True Positive Rate]

\addplot[color=black,mark=x]
plot coordinates {
(0,0.912466844)
(0.5,0.891246684)
(1,0.907161804)
(1.5,0.904509284)
(2,0.846153846)
};

\addlegendentry{WRF}

\addplot[color=black,mark=otimes]
plot coordinates {
(0,0.913660477)
(0.5,0.861007958)
(1,0.858488064)
(1.5,0.889124668)
(2,0.828514589)
};

\addlegendentry{CWRF}

\addplot[color=black,mark=triangle*]
plot coordinates {
(0,0.912466844)
(0.5,0.885941645)
(1,0.883289125)
(1.5,0.840848806)
(2,0.840848806)
};

\addlegendentry{IG}

\end{axis}
\end{tikzpicture}
\caption{True positive rate when varying $r$ on Spambase (incomplete knowledge)} 
\label{fig:spambase-r-tp}
\end{minipage}%
\hfill
\begin{minipage}[t]{0.49\linewidth}
\centering
   \begin{tikzpicture}[scale=0.7]
\begin{axis}[
ymin=0.5, ymax=10,
legend pos=south west,
line width=0.2mm,
mark size=2.5,
xlabel=$r$ value,
ylabel=Average Number of Modified Features]

\addplot[color=black,mark=x]
plot coordinates {
(0,4.100581395)
(0.5,4.043303571)
(1,5.627777778)
(1.5,7.851173021)
(2,7.593103448)
};

\addlegendentry{WRF}

\addplot[color=black,mark=otimes]
plot coordinates {
(0,4.322043912)
(0.5,5.129349285)
(1,6.323514978)
(1.5,8.869694214)
(2,8.703611752)
};

\addlegendentry{CWRF}

\addplot[color=black,mark=triangle*]
plot coordinates {
(0,4.100581395)
(0.5,4.900748503)
(1,4.95960961)
(1.5,5.055047319)
(2,4.892744479)
};

\addlegendentry{IG}

\end{axis}
\end{tikzpicture}
\caption{Average Number of Modified Features when varying $r$ on Spambase (incomplete knowledge)} 
\label{fig:spambase-r-av}
\end{minipage}\\
\begin{minipage}[t]{0.48\linewidth}
\centering
   \begin{tikzpicture}[scale=0.7]
\begin{axis}[
ymin=0.5, ymax=1,
legend pos=south west,
line width=0.2mm,
mark size=2.5,
xlabel=$r$ value,
ylabel=True Positive Rate]

\addplot[color=black,mark=x]
plot coordinates {
(0,0.955845166)
(1,0.955067108)
(2,0.937755301)
(3,0.936393698)
(4,0.934254036)
(5,0.932682809)
};
\addlegendentry{WRF}

\addplot[color=black,mark=otimes]
plot coordinates {
(0,0.956127213)
(1,0.949902743)
(2,0.937978992)
(3,0.935712896)
(4,0.930159502)
(5,0.936561743)
};
\addlegendentry{CWRF}

\addplot[color=black,mark=triangle*]
plot coordinates {
(0,0.955845166)
(1,0.955067108)
(2,0.937755301)
(3,0.936393698)
(4,0.934254036)
(5,0.939068275)
};
\addlegendentry{IG}

\end{axis}
\end{tikzpicture}
\caption{True positive rate when varying $r$ on Kyoto-Sample (incomplete knowledge)} 
\label{fig:kyoto-r-tp}
\end{minipage}%
\hfill
\begin{minipage}[t]{0.48\linewidth}
\centering
\begin{tikzpicture}[scale=0.7]
\begin{axis}[
ymin=0, ymax=5,
legend pos=south west,
line width=0.2mm,
mark size=2.5,
xlabel=$r$ value,
ylabel=Average Number of Modified Features]

\addplot[color=black,mark=x]
plot coordinates {
(0,2.830799756)
(1,2.631330618)
(2,3.643822167)
(3,3.949979227)
(4,4.02337428)
(5,3.521437474)
};
\addlegendentry{WRF}

\addplot[color=black,mark=otimes]
plot coordinates {
(0,2.616907845)
(1,2.813070792)
(2,3.710091369)
(3,3.898788615)
(4,3.969077118)
(5,3.699851445)
};
\addlegendentry{CWRF}

\addplot[color=black,mark=triangle*]
plot coordinates {
(0,2.830799756)
(1,2.705580448)
(2,3.701835719)
(3,3.97594516)
(4,3.99352488)
(5,3.797601163)
};
\addlegendentry{IG}

\end{axis}
\end{tikzpicture}
\caption{Average Number of Modified Features when varying $r$ on Kyoto-Sample (incomplete knowledge)} 
\label{fig:kyoto-r-av}
\end{minipage}\\
\vfill
\begin{minipage}[t]{0.48\linewidth}
\centering
\includegraphics[width=3.4in,height=2.2in]{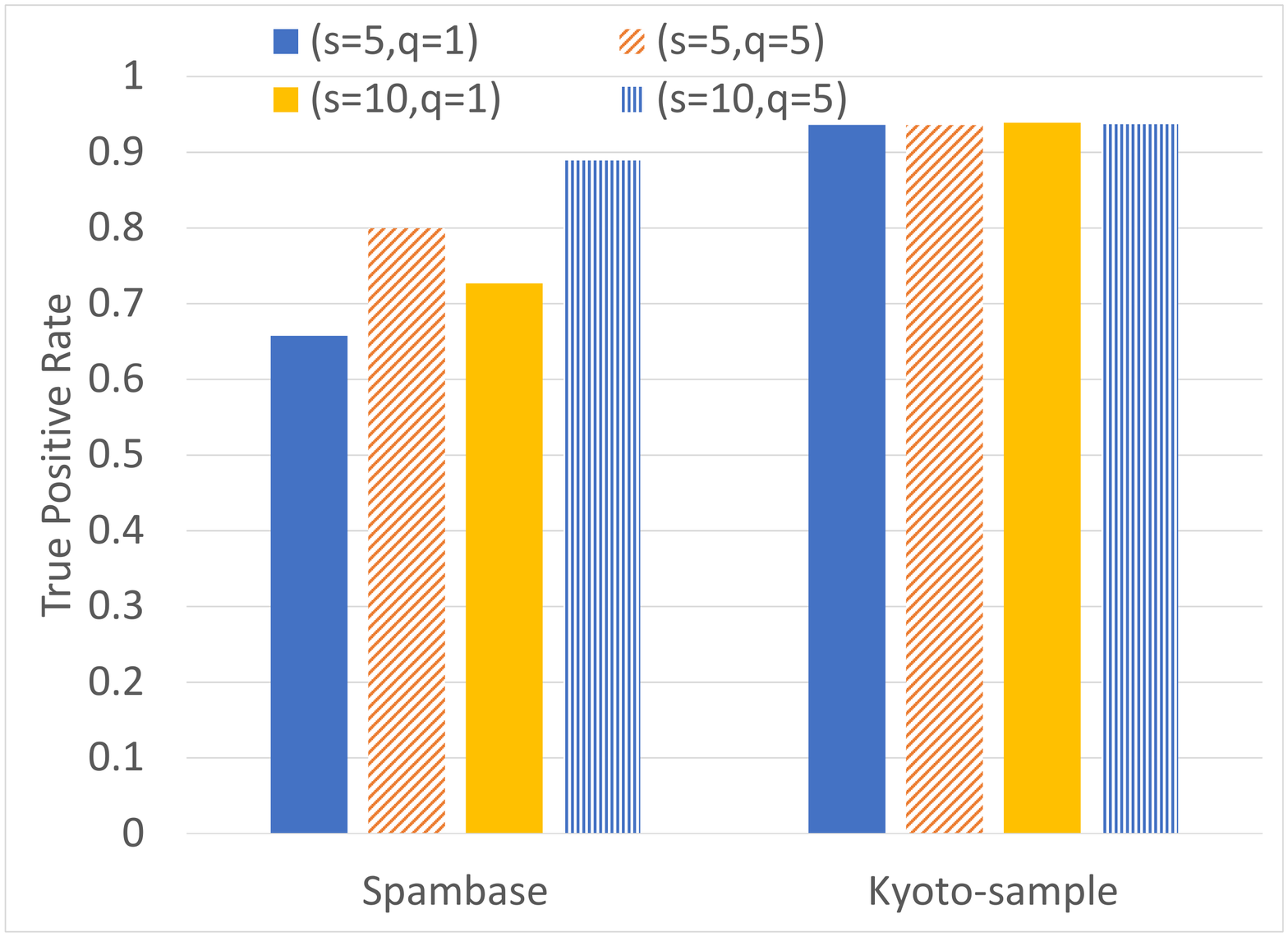}
\vspace{-0.5in}
\caption{True positive rate varying $s$ and $q$ (incomplete knowledge)}
\label{fig:s-tp}
\end{minipage}%
\hfill
\begin{minipage}[t]{0.48\linewidth}
\centering
\includegraphics[width=3.4in,height=2.2in]{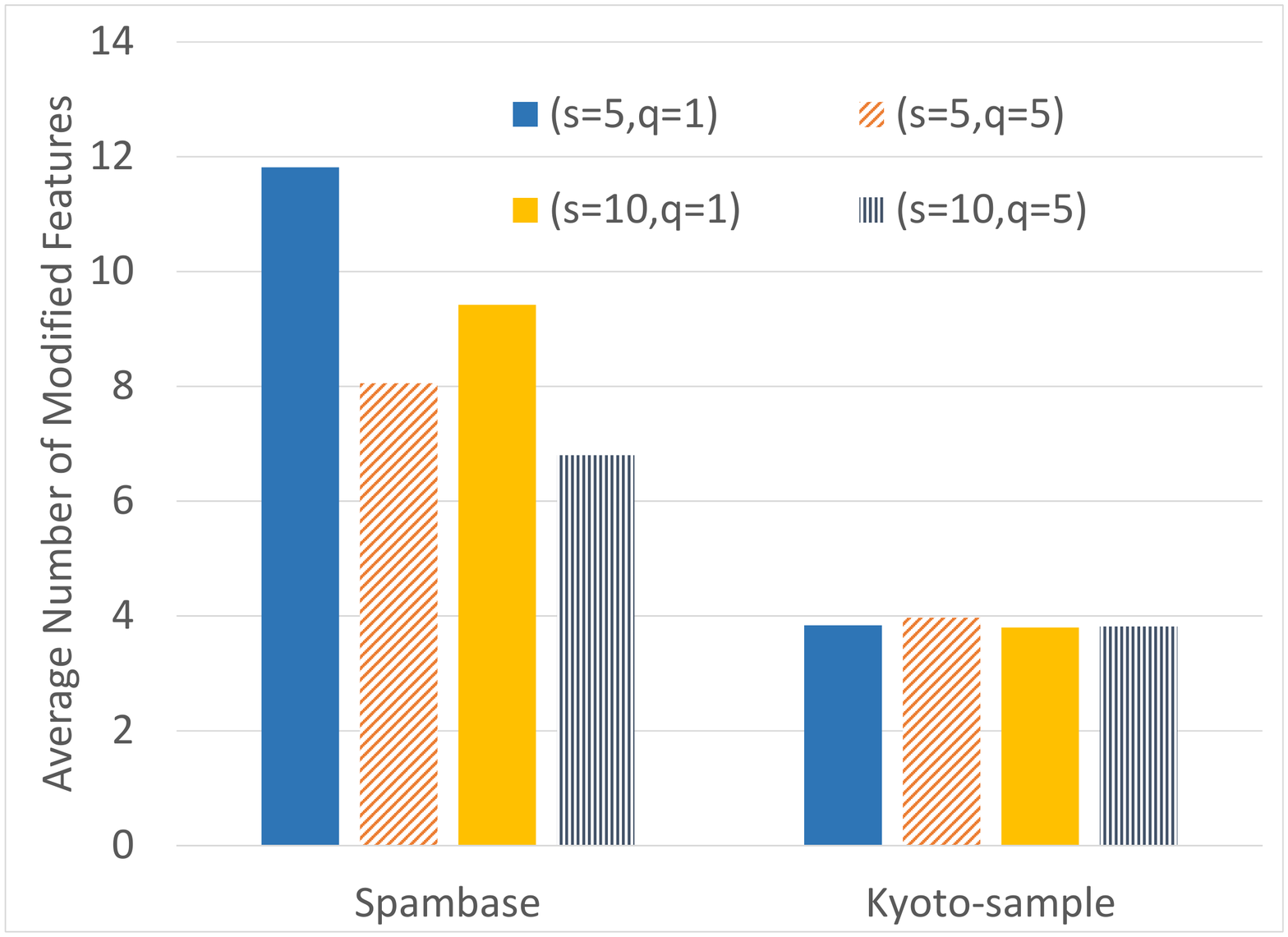}
\vspace{-0.5in}
\caption{Average Number of Modified Features varying $s$ and $q$ (incomplete knowledge)} 
\label{fig:s-av}
\end{minipage}\\
\begin{minipage}[t]{0.48\linewidth}
\centering
\includegraphics[width=3.4in,height=2.2in]{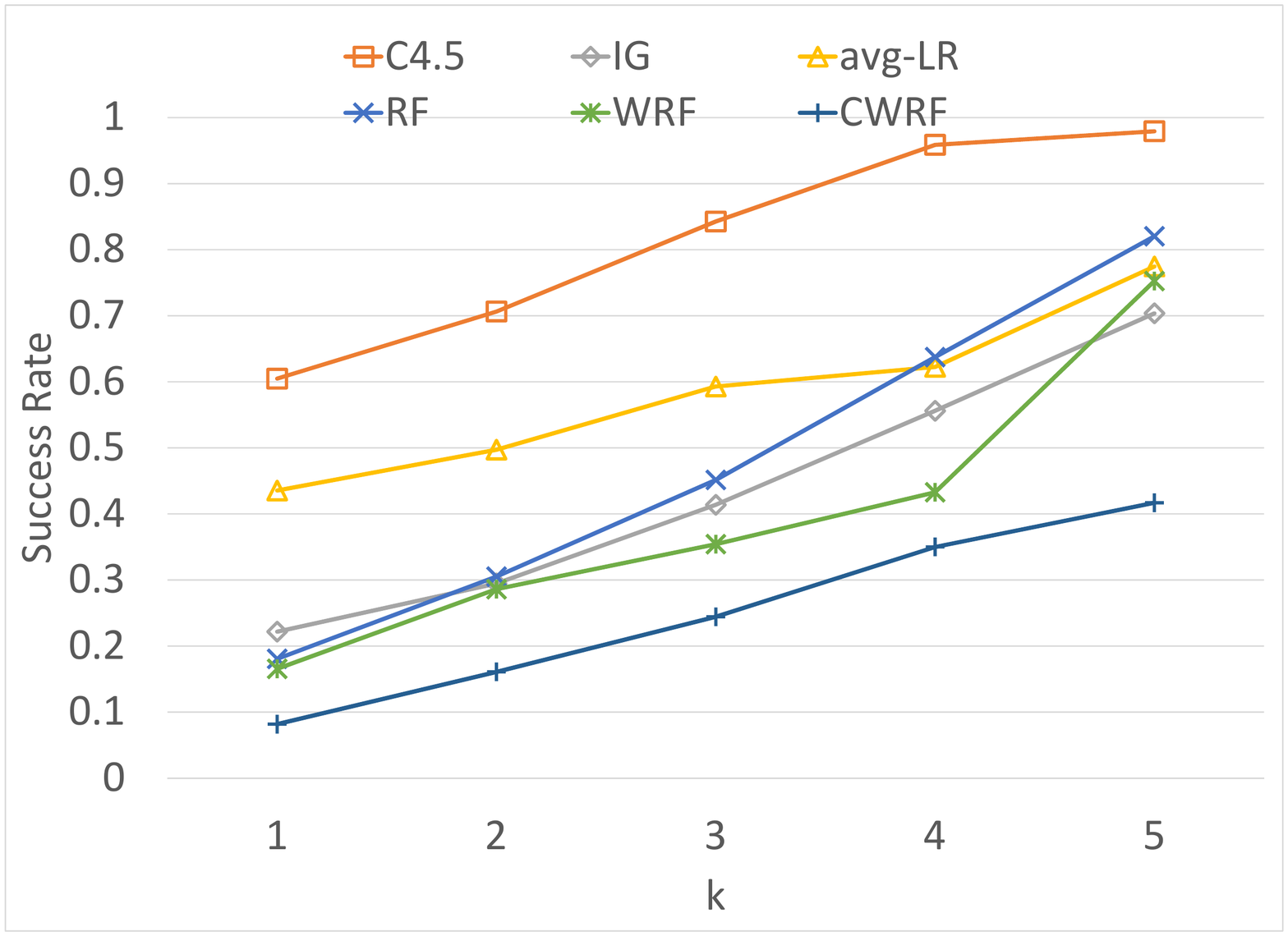}
\vspace{-0.5in}
\caption{Attackers' success rate when varying $k$ on Spambase (incomplete knowledge)}
\label{fig:spambase-k-succ}
\end{minipage}
\hfill
\begin{minipage}[t]{0.48\linewidth}
\centering
\includegraphics[width=3.4in,height=2.2in]{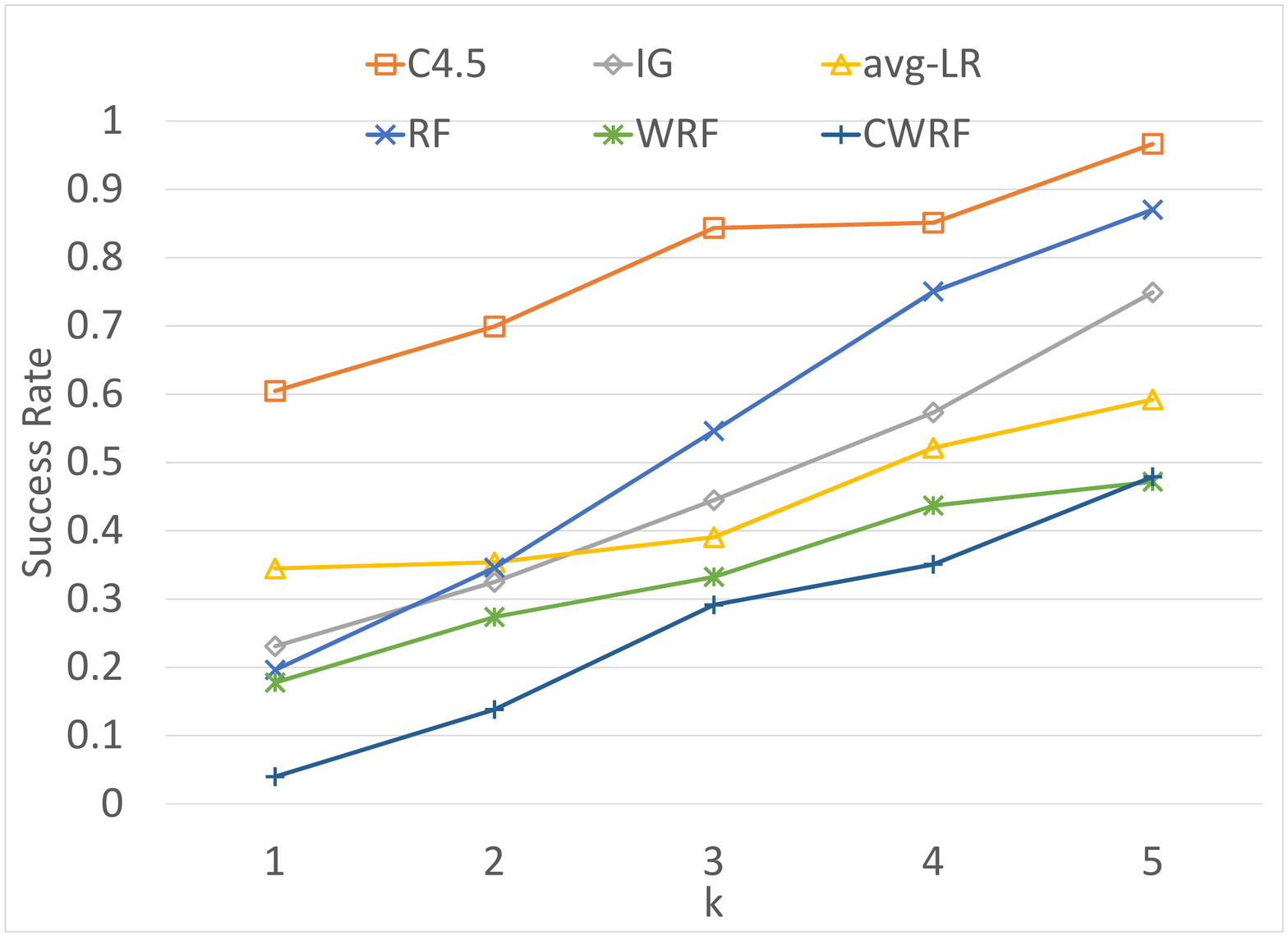}
\vspace{-0.5in}
\caption{Attackers' success rate when varying $k$ on Spambase (complete knowledge)} 
\label{fig:spambase-k-succ-comp}
\end{minipage}
\end{figure*}

We found that the optimal $r$ value is the same for both bounded and unbounded cost cases. 
Results for complete and incomplete knowledge also have similar trends so we only report results for incomplete knowledge and unbounded cost. 
As mentioned in Section \ref{sec:setup}, average number of modified features is used to measure robustness to evasion attacks in case of unbounded cost.
Figure \ref{fig:spambase-r-tp} and Figure \ref{fig:kyoto-r-tp} report true positive rate for Spambase and Kyoto-Sample with incomplete knowledge, respectively.
 Figure \ref{fig:spambase-r-av} and Figure \ref{fig:kyoto-r-av} report average number of modified features for each data set.
We reported results for WRF, CWRF, and IG because all these methods use $r$. 

The results showed that when $r$ is very small, all methods have high true positive rate (i.e., good mining quality). This is expected because more weights are put on mining quality. As $r$ increases, the average number of modified features increases and true positive rate decreases slightly. 

However as $r$ becomes too large, the average number of modified features becomes flat and even starts to decline. The reason is that for very large $r$ values, the weights for vulnerable features are so low such that other features that are previously not vulnerable may have higher weights and become vulnerable (i.e., used frequently in models). The optimal $r$ value is around 1.5 for Spambase. The optimal $r$ for Kyoto-Sample is higher (around 4.0), probably because this data set has fewer features and some of them are extremely vulnerable so a larger $r$ value is needed to achieve desired robustness level. 
In practice, users should start with small $r$ values and gradually increase it until average number of modified features or mining quality starts to drop significantly. 

We also observe that all three methods have similar true positive rates and their true positive rates are almost identical on Kyoto-Sample. In terms of average number of modified features, CWRF has the highest numbers on Spambase, meaning it is the most robust among these three methods. This is expected because CWRF adds randomness to both model training time and model application time. 
All three methods have similar average number of modified features for Kyoto-Sample, possibly due to fewer features in Kyoto-Sample.

We considered four combinations of $s$ and $q$: $s=5$ or $10$ (i.e., models are divided into 5 or 10 clusters) and $q=1$ or $5$ (1 or 5 models are selected per cluster). Again we found the optimal value for $s$ and $q$ is the same for both bounded and unbounded cost cases as well as for incomplete and complete knowledge cases so we only report results for unbounded cost and incomplete knowledge.
Figure \ref{fig:s-tp} reports true positive rate and Figure \ref{fig:s-av} reports average number of modified features for both data sets.

The results show that as $s$ and $q$ increases, true positive rate increases as well because more models are used in prediction. 
However the average number of modified features decreases in most cases as $s$ and $q$ increase because using more models means less uncertainty and less robustness. 
For Spambase, the decrease in true positive rate is quite significant as $s$ and $q$ decrease so the optimal setting is using 10 clusters and selecting 5 models per cluster. 
For Kyoto-Sample data set, true positive rate is not very sensitive to $s$ and $q$.  So we use 5 clusters and select 5 models from each cluster ($s=5,q=5$)
because it gives the highest average number of modified features.
In practice, users can start with large $s$ and $q$ and gradually decrease them until true positive rate starts to degrade significantly. 

\vspace{-0.1in}
\subsection{Comparison with Existing Methods}
\label{sec:result}

\begin{figure*}[!htb]
\begin{minipage}[t]{0.48\linewidth}
\centering
\includegraphics[width=3.4in,height=2.2in]{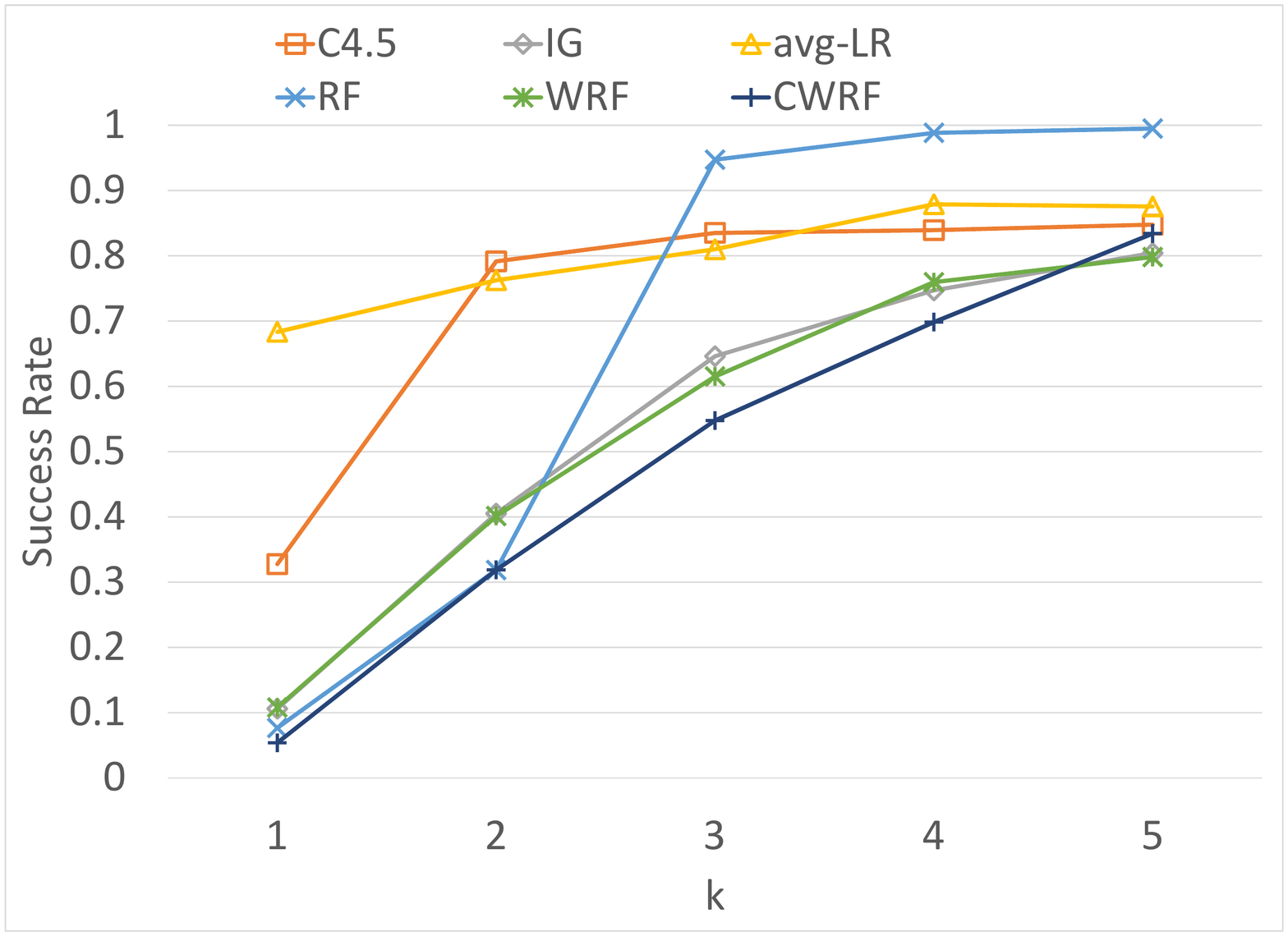}
\vspace{-0.4in}
\caption{Attackers' success rate when varying $k$ on Kyoto-Sample (incomplete knowledge)} 
\label{fig:kyoto-k-succ}
\end{minipage}
\hfill
\begin{minipage}[t]{0.48\linewidth}
\centering
\includegraphics[width=3.4in,height=2.2in]{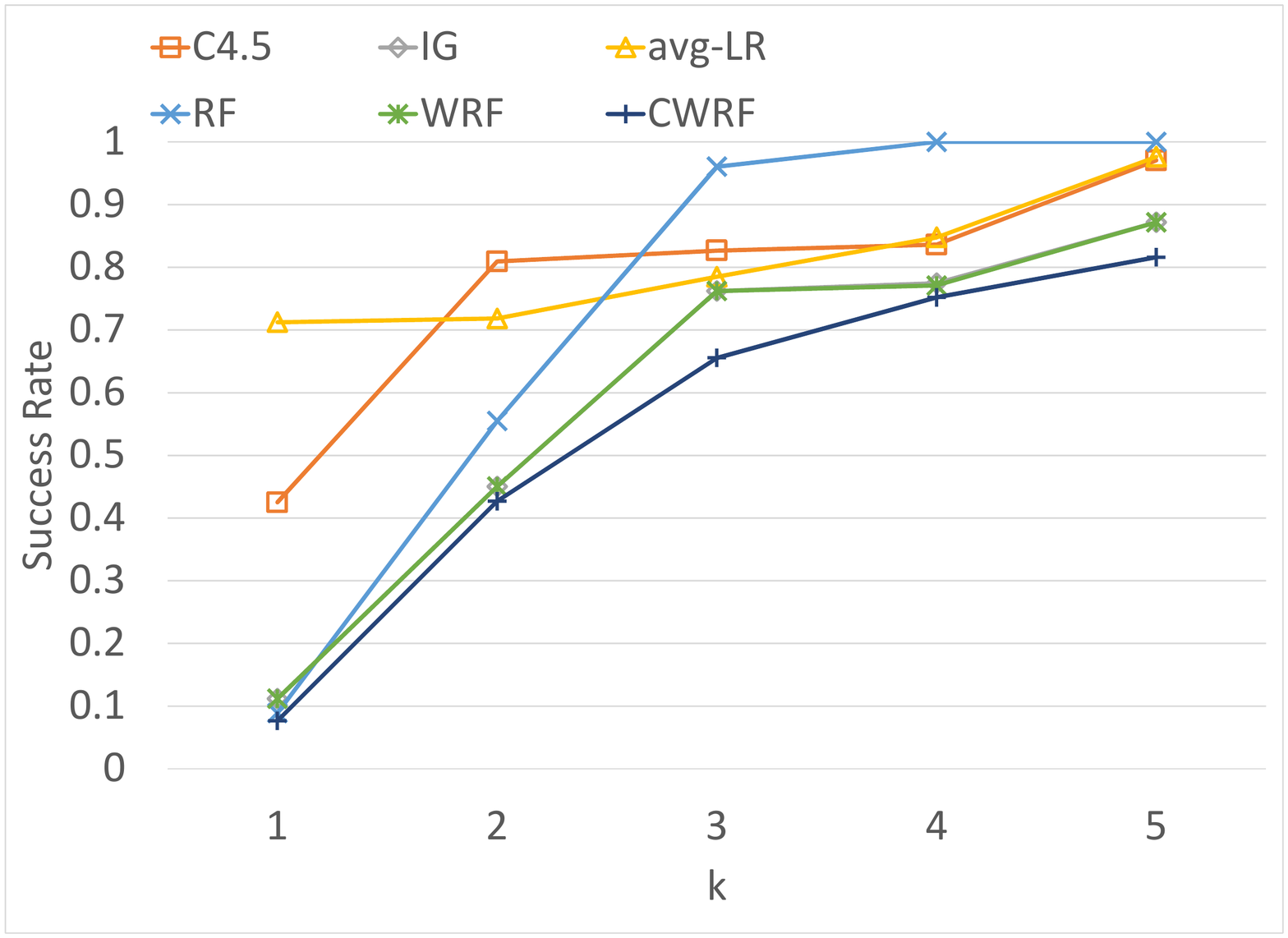}
\vspace{-0.4in}
\caption{Attackers' success rate when varying $k$ on Kyoto-Sample (complete knowledge)} 
\label{fig:kyoto-k-succ-comp}
\end{minipage}\\
\begin{minipage}[t]{0.48\linewidth}
\centering
\vspace{-0.1in}
\includegraphics[width=3.4in,height=2.2in]{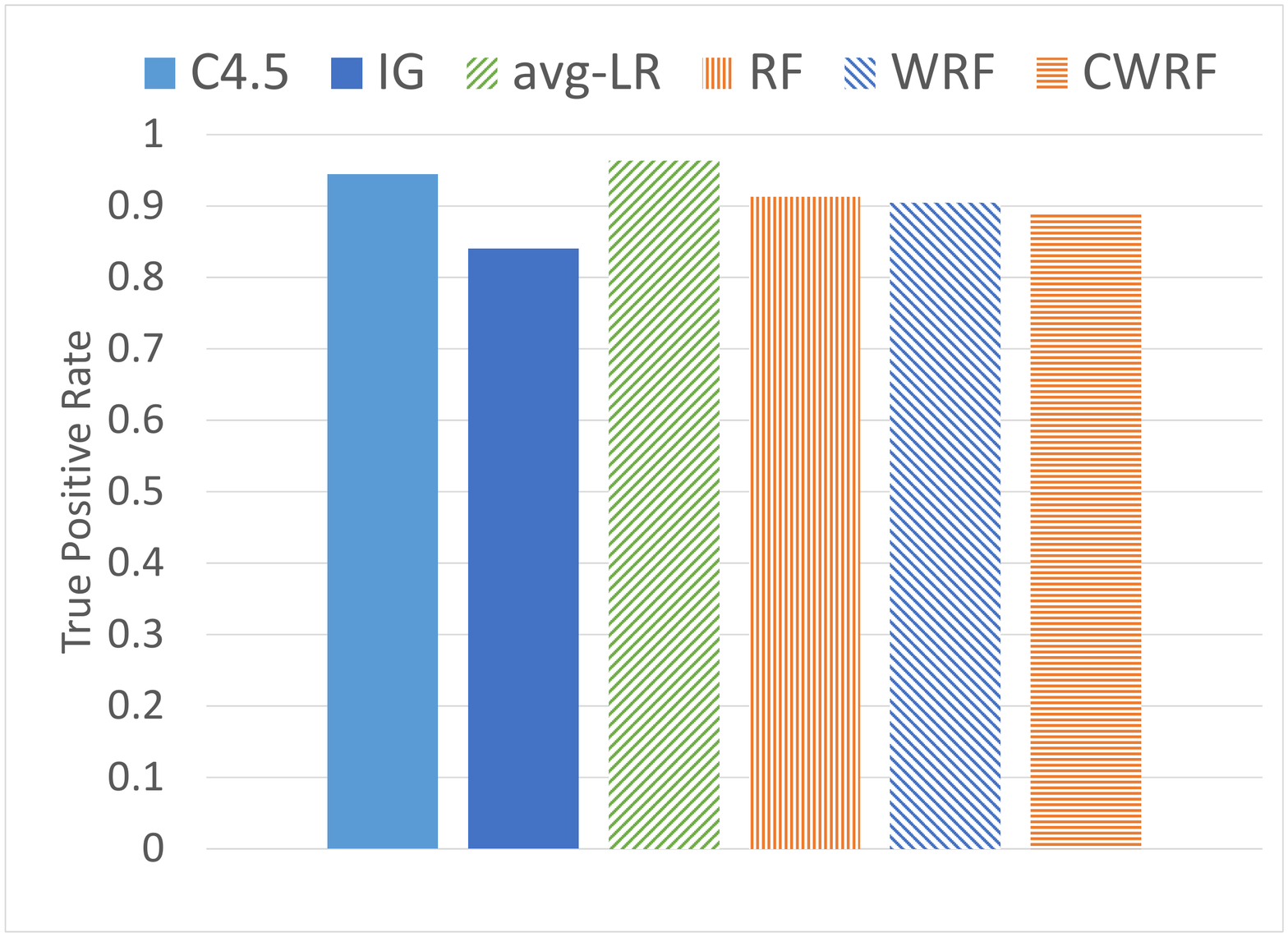}
\vspace{-0.4in}
\caption{True positive rate on Spambase}
\label{fig:spambase-tp}
\end{minipage}
\hfill
\begin{minipage}[t]{0.48\linewidth}
\centering
\vspace{-0.1in}
\includegraphics[width=3.4in,height=2.2in]{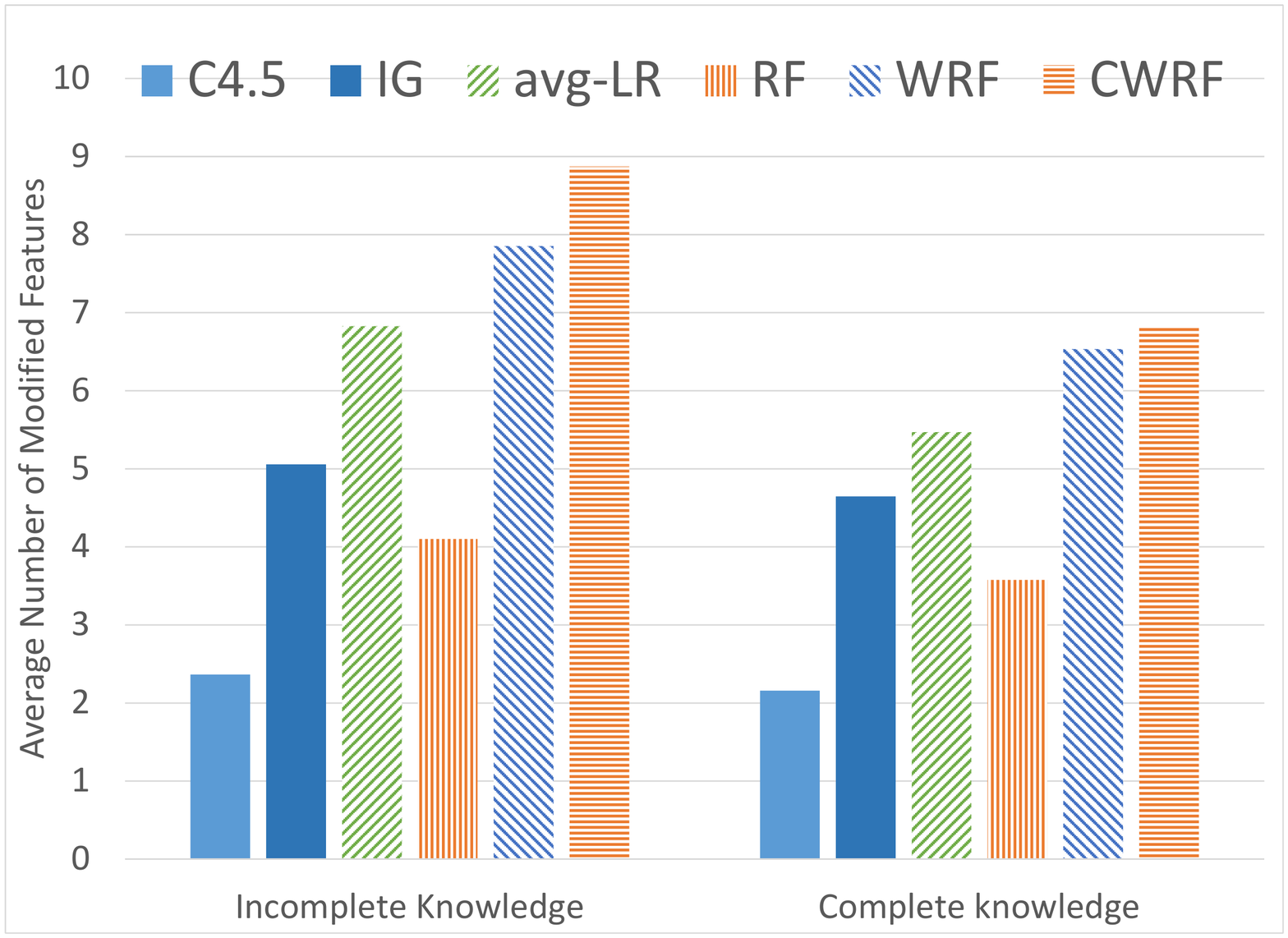}
\vspace{-0.4in}
\caption{Average Number of Modified Features  on Spambase} 
\label{fig:spambase-av}
\end{minipage}\\
\begin{minipage}[t]{0.48\linewidth}
\centering
\vspace{-0.1in}
\includegraphics[width=3.3in,height=2.2in]{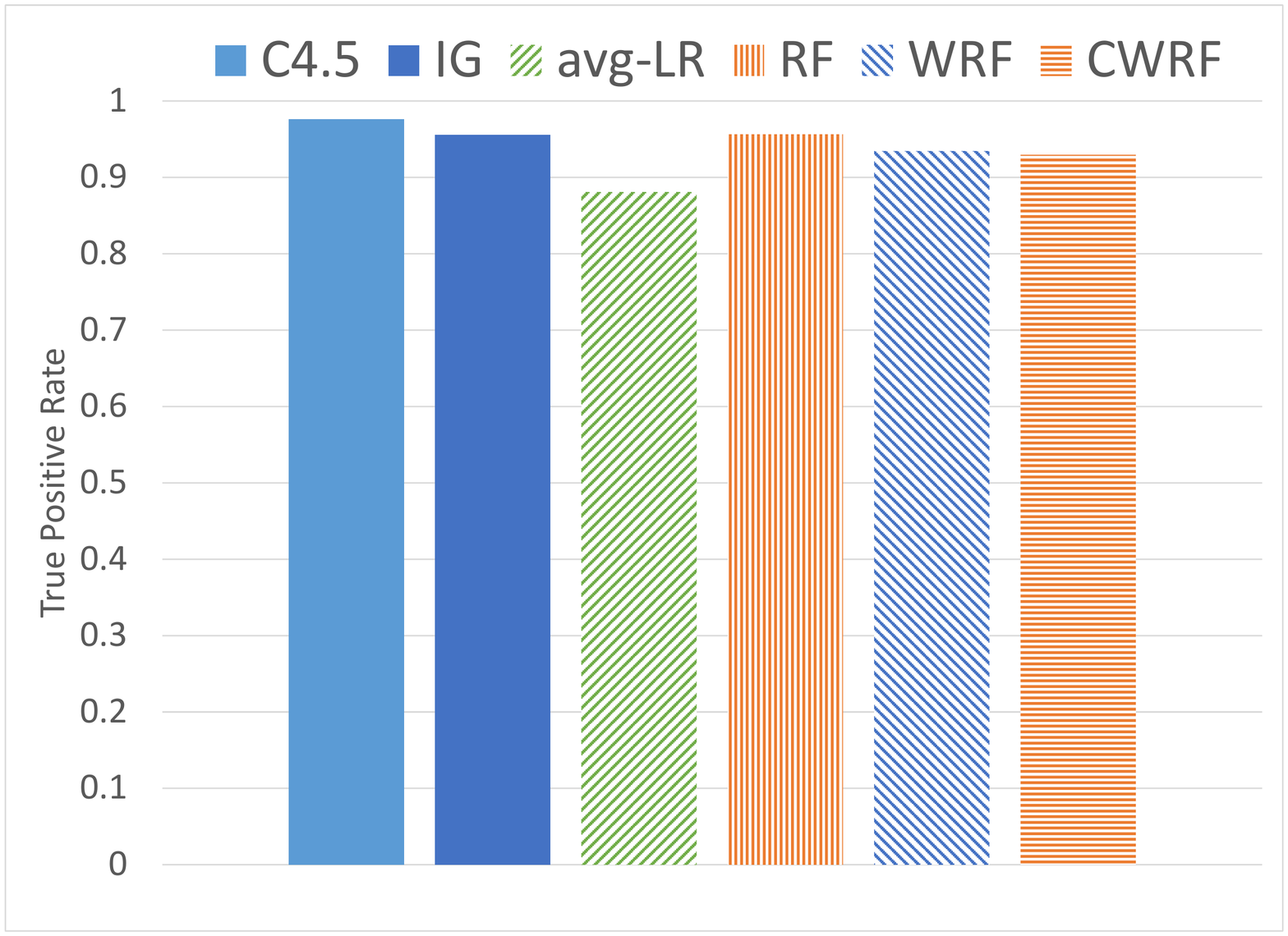}
\vspace{-0.3in}
\caption{True positive rate on Kyoto-Sample}
\label{fig:kyoto-tp}
\end{minipage}
\hfill
\begin{minipage}[t]{0.48\linewidth}
\centering
\vspace{-0.1in}
\includegraphics[width=3.3in,height=2.2in]{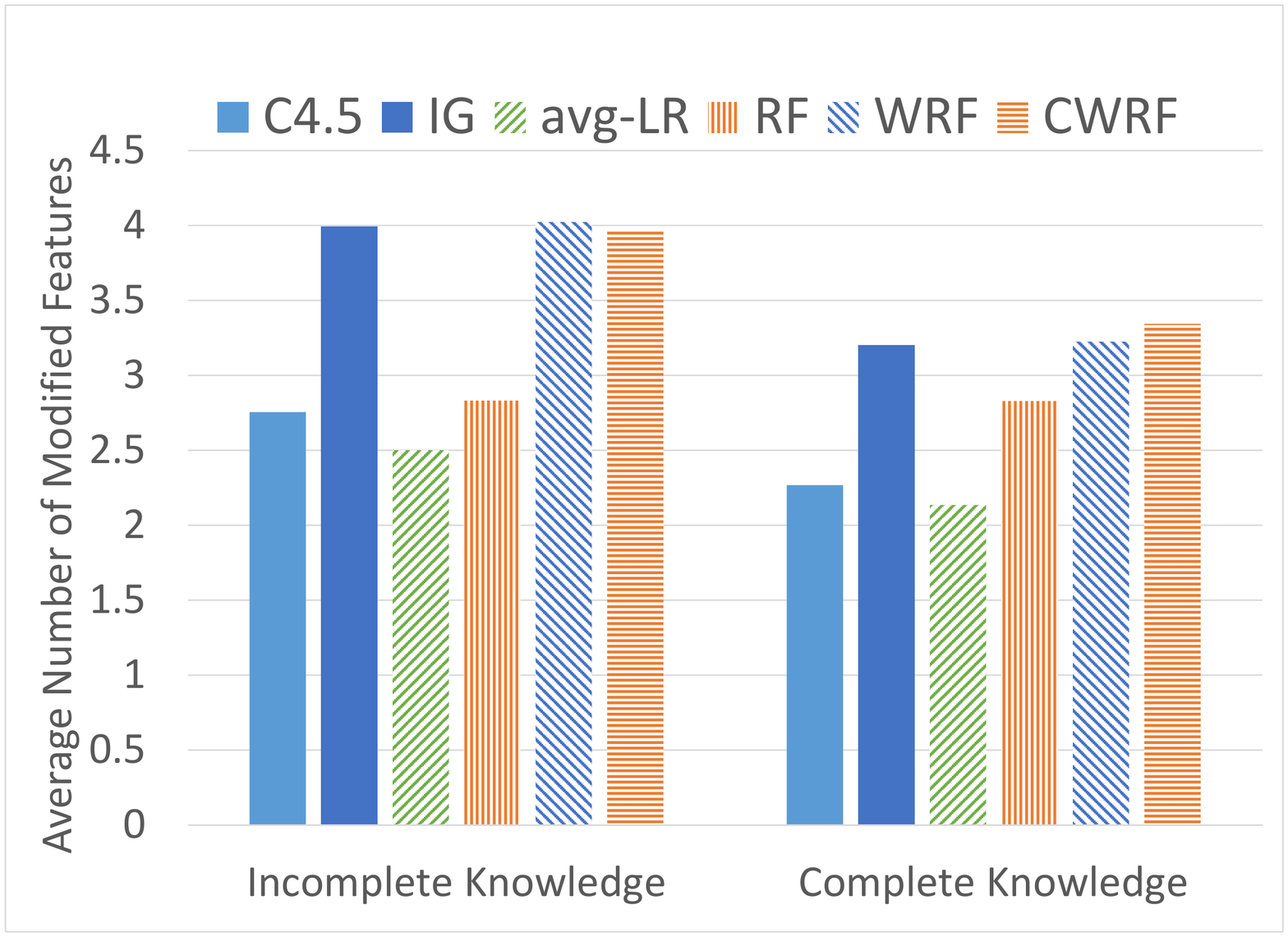}
\vspace{-0.3in}
\caption{Average Number of Modified Features on Kyoto-Sample} 
\label{fig:kyoto-av}
\end{minipage}\\
\begin{minipage}[t]{0.5\linewidth}
\centering
\vspace{-0.2in}
\includegraphics[width=3.3in,height=2.2in]{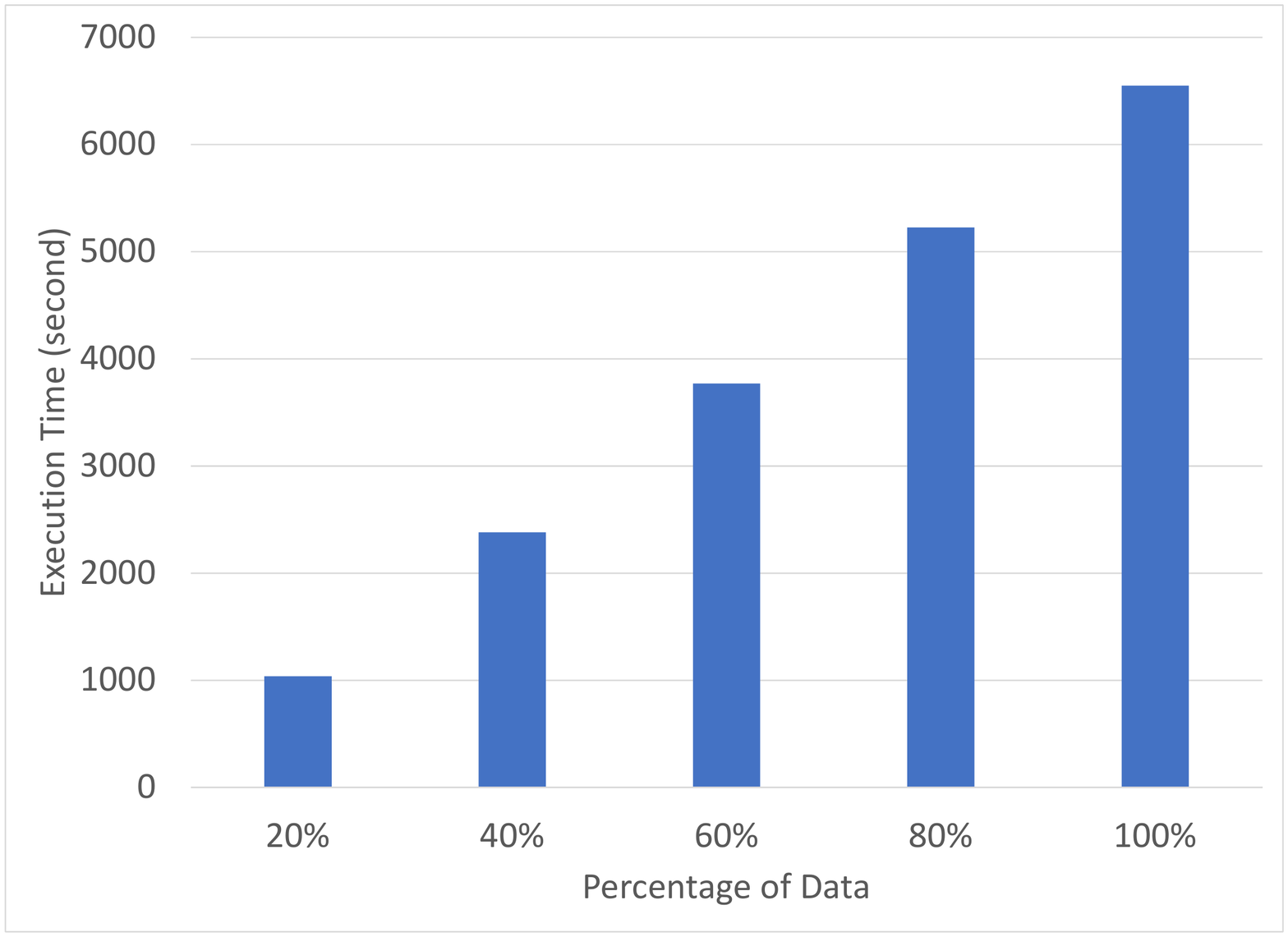}
\vspace{-0.3in}
\caption{Execution time of CWRF over fraction of Kyoto December 2015 data} 
\label{fig:scalability}
\end{minipage}
\end{figure*}

Once we set parameters for our algorithms (WRF and CWRF), we compare them to existing methods (RF, C4.5, avg-LR, and IG). 
We first consider bounded cost case and vary $k$ (number of features attackers can modify). 
Figure \ref{fig:spambase-k-succ} and Figure \ref{fig:spambase-k-succ-comp} report attackers' success rate for Spambase with incomplete and complete knowledge, respectively. 
Figure \ref{fig:kyoto-k-succ} and Figure \ref{fig:kyoto-k-succ-comp} report the results for Kyoto-Sample.
Note that true positive rate does not depend on attackers' actions, so true positive rate for bounded and unbounded cost cases are the same.
Figure \ref{fig:spambase-tp} and Figure \ref{fig:kyoto-tp} report true positive rate for Spambase and Kyoto-Sample, respectively.

As $k$ increases, attackers' success rate for all methods increases. This is expected because attackers can modify more features for larger $k$. 
For all methods attackers also have higher success rate with complete knowledge case than with incomplete knowledge. This is also expected because 
attackers can launch more effective attacks if they know more. 
Due to small number of features (only 14 in total), all methods have higher success rate on Kyoto-Sample data set, especially for large $k$ values. 

The results show that CWRF has the lowest success rate for attackers for most cases, meaning it is the most robust method. 
The improvement on robustness is quite significant in most cases. 
For example, on Spambase attackers' success rate is 64\% when $k=4$ using random forest. 
Using CWRF attackers success rate is reduced to 35\%. 
On Kyoto-Sample attackers' success rate is 95\% when $k=3$ using random forest and using CWRF the success rate is reduced to 55\%. 

Figure \ref{fig:spambase-tp} and Figure \ref{fig:kyoto-tp} also show that
true positive rate for CWRF and WRF is quite close to that of random forest. So the improvement of robustness does not come at significant cost of 
detection rate. 

The results also show that WRF has the second lowest success rate for attackers and beats IG in most cases except for complete knowledge case on Kyoto-Sample where IG and WRF have the same performance. 
WRF uses differential ratio and IG uses information gain at model training time. This verifies the
observation in  Section \ref{sec:weight} that information gain is often inferior to differential privacy in measuring vulnerability. 
CWRF is better than WRF because CWRF also injects randomness into model application time. 

RF has lower success rate for attackers than C4.5 on Spambase. On Kyoto-Sample the results are mixed as RF is better for small $k$ values but worse than C4.5 for large $k$ values. This shows that adding randomness to model building time helps robustness in general but more can be done as demonstrated by the results of CWRF. 

The results for avg-LR are always worse than CWRF and WRF. Avg-LR does use randomness at model building time but it neither uses randomness at model application time nor optimizes tradeoff between mining quality and robustness (it still uses conventional logistic regression when building each model). 

For unbounded cost case, Figure \ref{fig:spambase-av} and Figure \ref{fig:kyoto-av} report average number of modified features for Spambase and Kyoto-Sample, respectively. CWRF has the highest average number of modified features (i.e., highest attackers' effort) for Spambase, followed by WRF, avg-LR, and IG.
The improvement of CWRF over RF is quite significant. For example, in incomplete knowledge case attackers need to modify on average 4.1 features to evade detection from RF but need to modify 8.9 features to evade detection from CWRF. 

For Kyoto-Sample, IG, CWRF, and WRF have the highest average number of modified features. The difference between these three methods is quite small.
One possible reason is that Kyoto-Sample has fewer features so all models have a lot of overlapping features and clustering becomes less effective. Still the results of CWRF are significantly better than that of random forest. For example, in incomplete knowledge case attackers need to modify on average 2.8 features to evade detection from RF but need to 
modify 4.0 features (there are only 14 features in total) to evade detection from CWRF.


\noindent
{\bf Scalability:} We extracted 20\%, 40\%, 60\%, 80\% as well as the full set of Kyoto data set collected in December 2015 (with 7,565,246 million instances in total) to test the scalability of CWRF. 
Figure \ref{fig:scalability} reports the execution time of CWRF.
The results show that CWRF scales linearly with number of rows. We also found that the execution time is dominated by model building time (clustering time is less than 10 seconds). 
As future work we will investigate how to further improve efficiency of our methods.

\vspace{-0.1in}
\section{Conclusion}
\label{sec:conclusion}
This paper proposes an approach to use randomization to improve robustness of machine learning models in cyber security applications. 
Our approach injects randomness into both model training time and model application time and carefully balances robustness and mining quality.
We applied our approach to random forest and experiments on an email spam data set and a network intrusion detection data set
show our approach significantly improves robustness of random forest models without sacrificing much mining quality. 
We also discuss some ideas of how to extend our approach to other mining methods. 
For future work it will be interesting to study effectiveness of these extensions. 
We also made a number of simplified assumptions in our experiments (e.g., uniform cost of modifying each feature). In future work we will study results in more realistic settings, e.g., with different cost of modifying each feature and some features cannot be modified (e.g., the part of malware that carries out the attack).

\bibliographystyle{ACM-Reference-Format}
\bibliography{privacy}
\end{document}